\documentclass[aps,nofootinbib,superscriptaddress]{revtex4}

\usepackage[normalem]{ulem}

\usepackage{amsmath}
\usepackage{amsfonts,amssymb,amsthm, bbm,braket}
\usepackage{graphicx}   
\usepackage{subfigure}  
\usepackage{amsbsy} 
\usepackage[bold]{hhtensor}  

\usepackage{natbib}
\usepackage{algorithm}
\usepackage{algpseudocode}

\usepackage{multirow}

\newcommand{\expect}{\mathbb{E}}
\newcommand{\diag}{\operatorname{diag}}

\newcommand\T{\operatorname{T}}
\newcommand\dif{\mathrm{d}}
\newcommand\Var{\operatorname{Var}}
\newcommand\Cov{\operatorname{Cov}}
\newcommand\Vol{\operatorname{Vol}}
\newcommand\cdf{\operatorname{cdf}}
\newcommand\erf{\operatorname{erf}}
\newcommand\Sym{\operatorname{Sym}}

\newcommand\Tr{\mathrm{Tr}}

\newcommand\id{\mathbbm 1}

\newcommand\doi[1]{\href{http://dx.doi.org/#1}{doi:#1}}

\global\def \arxivmode {}

\ifx \arxivmode \undefined
\else
  
  \newcommand\arxivonly[1]{#1}
  \newcommand\prlonly[1]{}
\fi

\ifx \prlmode \undefined
\else
  
  \newcommand\arxivonly[1]{}
  \newcommand\prlonly[1]{#1}
\fi

\usepackage{hyperref} 

\begin{document}

\title{Robust Online Hamiltonian Learning}

\author{Christopher E. Granade\footnote{Corresponding author (cgranade@cgranade.com).}}
\affiliation{
Institute for Quantum Computing,
University of Waterloo,
Waterloo, Ontario, Canada}
\affiliation{
Department of Physics,
University of Waterloo,
Waterloo, Ontario, Canada}

\author{Christopher Ferrie}
\affiliation{
Institute for Quantum Computing,
University of Waterloo,
Waterloo, Ontario, Canada}
\affiliation{
Department of Applied Mathematics,
University of Waterloo,
Waterloo, Ontario, Canada}

\author{Nathan Wiebe}
\affiliation{
Institute for Quantum Computing,
University of Waterloo,
Waterloo, Ontario, Canada}

\author{D.G. Cory}
\affiliation{
Institute for Quantum Computing,
University of Waterloo,
Waterloo, Ontario, Canada}
\affiliation{
Department of Chemistry,
University of Waterloo,
Waterloo, Ontario, Canada}
\affiliation{
Perimeter Institute for Theoretical Physics,
Waterloo, Ontario, Canada}

\date{\today}


\begin{abstract}
In this work we combine two distinct machine learning methodologies, sequential Monte Carlo and Bayesian experimental design, and apply them to the problem of inferring the dynamical parameters of a quantum system.  We design the algorithm with practicality in mind by including parameters that control trade-offs between the requirements on computational and experimental resources.  The algorithm can be implemented \emph{online} (during experimental data collection), avoiding the need for storage and post-processing.  Most importantly, our algorithm is capable of learning Hamiltonian parameters even when the parameters change from experiment-to-experiment, and also when additional noise processes are present and unknown.  The algorithm also numerically estimates the Cramer-Rao lower bound, certifying its own performance.  
\end{abstract}

\maketitle

\section{Introduction}
Building a large scale quantum information processor is a significant challenge. First, we
require an accurate characterization of the dynamics experienced by the device to allow for
the application of error correcting codes and other tools for implementing useful quantum
algorithms. Characterization is carried out through the statistical estimation of parameters describing the quantum states and processes involved (also called \emph{tomography}) \cite{2004Quantum}.  This characterization problem is especially timely, since 
quantum simulation experiments are approaching a complexity where classical computers
are unable to simulate their evolution~\cite{LHM+11,GLK+11,KCK+10}.  In cases where so called analog quantum simulation
is applied, the validity of the simulation directly depends on the accuracy with which the Hamiltonian of the quantum
simulation conforms to the dynamical model that the experimenter believes describes the simulator. At present,
such simulators are certified by comparing their outputs to those expected from a classical--computer simulation~\cite{GLK+11,KCK+10}.  This means that new methods for Hamiltonian characterization will be vital for
certifying the next generation of quantum simulators.

In quantum tomography, the usual scenario discussed has been \emph{full} quantum process estimation -- there really is a ``black
box'' that the experimenter knows nothing about. While conservative, this is highly unlikely in practice because physics typically
gives some insight into the form of the Hamiltonian which gives rise to
the process. This additional knowledge reduces the number of parameters of the
process and processes for learning these parameters are sometimes called \emph{partial} process estimation or partial tomography~\cite{BPP08,BPP09,MR09,BNW+09,Flammia2011Direct,daSilva2011Practical}.
If our goal is to build a quantum information processing device, we must consider also an
additional complication: a characterization of a process at a ``snap-shot'' in time is not
nearly as useful as a characterization of the dynamics a quantum system undergoes. The evolution of closed systems is
given by a Hamiltonian operator, and hence this process is usually called Hamiltonian estimation in such cases.

One way to adapt the above schemes to Hamiltonian estimation is by stroboscopically
estimating snap-shots of the process at fixed times and then use various algorithms to invert
these to find the Hamiltonian via post--processing\footnote{See \cite{Oi2012Quantum} and references therein.}.
 The key additional freedom we consider has received little attention to date, and is that the controls after some number of measurements can depend on the outcomes of those previous measurements.  Hence, we call our derived strategies \emph{adaptive} or \emph{online}.  Under its very broad definition, our method can be called \emph{machine learning};  however, a more descriptive name is \emph{sequential Monte Carlo Bayesian experimental design}.  The marriage of sequential Monte Carlo methods \cite{Doucet2009Tutorial} and Bayesian experimental design methods \cite{Loredo2004Bayesian} has been considered very recently in a wide variety of classical contexts \cite{Kuck2006SMC,Scarpa2007Bayesian,Cavagnaro2010Adaptive,Kantas2010Simulationbased,Huan2011Simulationbased} and also for measurement adaptive quantum state tomography \cite{Huszar2012Adaptive}\footnote{It is interesting to note, however, that online experimental design may be unnecessary (asymptotically requiring only one adaptive step) for the particular state
tomography problem considered there \cite{Bagan2006Separable}.}.  Other machine learning ideas have also been generalized to the quantum domain \cite{Servedio2004Equivalences,Aimeur2006Machine,Aaronson2007Learnability,Hentschel2010Machine,Pudenz2011Quantum,Hentschel2011Efficient,Sergeevich2012Optimizing}. In this paper we present such an algorithm for learning dynamical parameters of quantum mechanical systems.

We make this learning process tractable by utilizing information about a system, rather than starting from worst-case assumptions such as those made in traditional quantum process and state tomography. We often have in practice knowledge about the dynamical model that describes a system of interest, and wish to improve that knowledge by estimating specific model parameters.  Thus, practical Hamiltonian finding can often be achieved via a suitable parameterization of the Hamiltonian, $H(x_1, \dots, x_d)$, reducing the problem to estimating the vector of parameters $\vec{x} = (x_1, \dots, x_d)$.  The task we consider is the design of experiments for the purpose of deducing these parameters in the smallest number of experiments possible.    Our algorithm also provides a \emph{region estimation} for the Hamiltonian parameters that encloses some fixed volume of parameter space in which the mean or the variance of the Hamiltonian parameters are expected to be found with
high--probability.  We also generalize this concept to allow the algorithm to learn \emph{hyperparameters}, which describe the distribution of the Hamiltonian parameters in cases where the parameters randomly vary between experiments.

This paper is organized as follows.  In section \ref{sec:BED} we review the formalism of Bayesian experimental design.  Next, we discuss the statistical metrics we employ and the Cramer-Rao bound in section \ref{sec:stats_stuff}.  Section \ref{sec:SMC} introduces the sequential Monte Carlo algorithm.  The test cases we use for the numerical experiments are described in section \ref{sec:test cases}.  In sections \ref{sec:region-est} and \ref{sec:hyperparam-est}, we discuss the application of our algorithm to region estimation and hyperparameter estimation, respectively. We explore the implications of the numerical benchmarking results in section \ref{sec:benchmarking}.

\section{Experimental Design Formalism}
\label{sec:BED}

The key element which interfaces quantum theory and machine learning is the equivalence of the \emph{Born rule} from quantum theory and the \emph{likelihood function} from statistics \cite{caves_quantum_1986}.  Each quantum mechanical problem specification produces a probability distribution $\Pr(d_k|\vec{x};c_k)$, where $d_k$ is the data obtained and $c_k$ are the experimental designs (or \emph{controls}) chosen for measurement $k$, and where $\vec{x}$ is a vector parameterizing the system of interest.

Suppose we have performed experiments with control settings $C:=\{c_1,c_2,\ldots,c_N\}$ and obtained data $D:=\{d_1,d_2,\ldots,d_N\}$.  The model specifies the likelihood function
\[
\Pr(D|\vec{x};C) = \prod_{k=1}^N \Pr(d_k|\vec{x};c_k).
\]
However, we are ultimately interested in $\Pr(\vec{x}|D;C)$, the probability distribution of the model parameters $\vec{x}$ given the experimental data.  We achieve this using use Bayes' rule:
\[
\Pr(\vec{x}|D;C)=\frac{\Pr(D|\vec{x};C)\Pr(\vec{x})}{\Pr(D|C)},
\]
where $\Pr(\vec{x})$ is the \emph{prior}, which encodes any \emph{a priori} knowledge of the model parameters.  The final term $\Pr(D|C)$ can simply be thought as a normalization factor.  We refer to this update proceedure as batch processing because
the experimental controls $C$ do not depend on the observed data and hence the processing for the experiment design can
be done offline (meaning after all data is collected).

An alternative to batch processing of given data is to adaptively choose the controls for the next experiment given the data from the past experiment.  This idea can be formalized in various ways -- the most natural for our purposes being called \emph{Bayesian experimental design} \cite{Loredo2004Bayesian}.  For this we conceive of possible future data $d_{N+1}$ obtained from a, possibly different, set of experimental controls $c_{N+1}$.  The probability of obtaining this data can be computed from the distributions at hand via marginalizing over model parameters
\[
\Pr(d_{N+1}|D; c_{N+1}, C)=\int  \Pr(d_{N+1}|\vec{x};c_{N+1})\Pr(\vec{x}|D;C) d\vec{x}.
\]
Note, in the remainder we will use the following abbreviated notation for expectation values:
\begin{equation}
\Pr(d_{N+1}|D; c_{N+1}, C)=\mathbb{E}_{\vec{x}|D;C}[\Pr(d_{N+1}|\vec{x};c_{N+1})],\label{eq:sec2:expectation}
\end{equation}
where the subscript on $\mathbb{E}$ denotes the variable for the expectation to be taken over.

The expectation value in~\eqref{eq:sec2:expectation} can be used to inform the algorithm about the choices of experimental
parameters that are more useful than others.  This usefulness is quantified, for a given choice of a \emph{utility function} $U(D;C)$, by the  expected \emph{utility} of an experiment
\[
U(c_{N+1})=\mathbb{E}_{d_{N+1}|D; c_{N+1}, C}[U(d_{N+1};c_{N+1})],
\]
where $U(d_{N+1};c_{N+1})$ is the utility we would derive if experiment $c_{N+1}$ yielded result $d_{N+1}$.  The
choice of the utility function is motivated by the figure of merit that we want to optimize.  We will consider two canonical choices: \emph{information gain} and the negative \emph{variance} and discuss
them in detail in the subsequent section.

\section{Utility Functions and the Cramer-Rao Lower Bound\label{sec:stats_stuff}}

Given a set of observed outcomes, the choice of subsequent experimental parameters that informs us most about the model parameters is given by the \emph{utility function}.
A generally well motivated measure of utility for scientific inference is \emph{information gain} \cite{Lindley1956On,SBT11}.  In information theory, information is measured by the entropy
\[
U(d_{N+1};c_{N+1})=\mathbb{E}_{\vec{x}|d_{N+1},D;c_{N+1},C}[\log\Pr(\vec{x}|d_{N+1},D;c_{N+1},C)].
\]
Maximizing the expected value of this utility function is equivalent to minimizing the expected entropy in the posterior distribution, $\Pr(\vec{x}|d_{N+1},D;c_{N+1},C)$.  We also test or method with a utility function that minimizes the expected variance in $\Pr(\vec{x}|d_{N+1},D;c_{N+1},C)$.  We show that this choice is optimal for minimizing the the mean squared error of the protocol.

\begin{figure}\centering
  \includegraphics[height=.6\columnwidth]{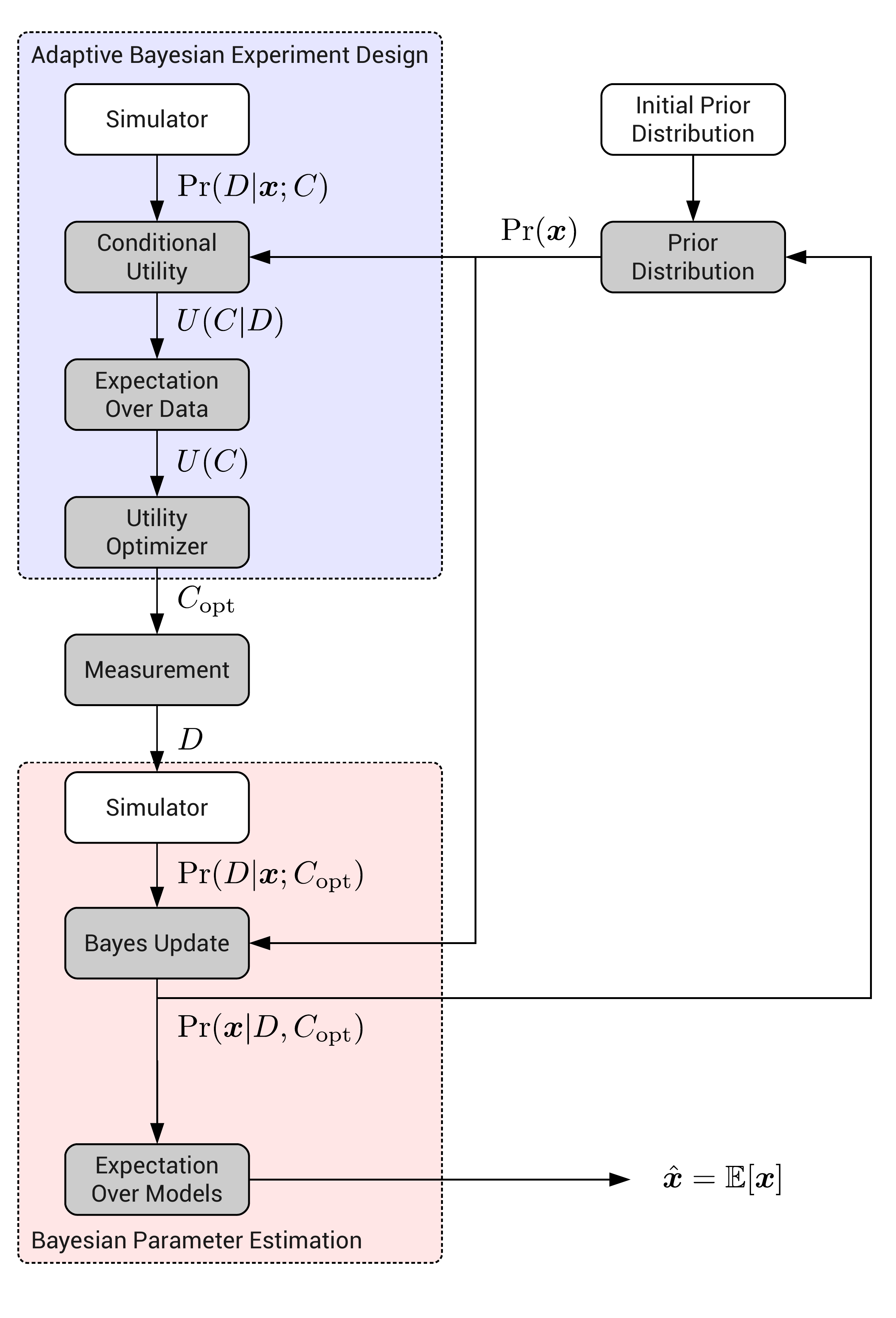}
  \caption{\label{fig:algorithm-block-diagram-crop} Overview of algorithm the combined experimental design and parameter estimation algorithm, showing the posterior distribution as a feedback into the next iteration. Blocks with a white background indicate model-dependent steps.}
\end{figure}

An \emph{estimator} is a function $\hat{\vec{x}}$ that takes a set of observed data $D$ collected from a set of experiments with controls $C$ and produces an estimate for the unknown parameters $\vec{x}$. Here, we evaluate the quality of an estimator $\hat{\vec{x}}$ by using a generalization of the \emph{squared error loss} called the \emph{quadratic loss} as our figure of merit. The quadratic loss is defined for a vector of parameters $\vec{x}$, data $D$ and experiment designs $C$, as
\begin{equation}
\label{eq:mse-loss}
L_{\matr{Q}}(\vec{x},\hat{\vec{x}}(D,C)) = \left(\vec{x}-\hat{\vec{x}}(D, C)\right)^\text{T}\matr{Q}\left(\vec{x}-\hat{\vec{x}}(D,C)\right),
\end{equation}
where $\matr{Q}$ is a positive definite matrix on the space of unknown parameters that defines the relative scale between the various parameters of interest.
The quadratic loss function is useful to us in that it is computationally inexpensive to calculate and may be analyzed by well-known statistical techniques. In particular, the Cramer-Rao bound can be used to lower-bound the mean quadratic loss incurred by an estimator, under the hypothesis of a given true model $\vec{x}$ \cite{Lehmann1998Theory}.

Following a decision theoretic methodology \cite{Berger1985Statistical}, the \emph{risk} of an estimator given a set of experiment designs $C$ is its expected performance over all possible outcomes $D$ with respect to the loss function:
\[
R({\vec{x}},\hat{\vec{x}};\ C) = \mathbb{E}_{D|\vec{x};C}[L(\vec{x},\hat{\vec{x}}(D;C))].
\]
The Bayes risk is the average of this quantity with respect to a prior distribution on $\vec{x}$ (denoted $\pi$) and is given explicitly by
\begin{align*}
r(\pi;C) &= \mathbb{E}_{\vec{x}}[R({\vec{x}},\hat{\vec{x}};\ C)]\nonumber\\
&=\int \pi(\vec{x}) R({\vec{x}},\hat{\vec{x}};\ C)\dif\vec{x}.
\end{align*}
where $\hat{\vec{x}}$ is assumed to be a \emph{Bayes estimator}, which means it is the one which minimizes the Bayes risk.  When the loss function is taken to be squared error (in the single parameter case) or the quadratic loss (in the multi-parameter case), the Bayes risk is more familiarly known as \emph{mean squared error} (MSE).

For quadratic loss (and many others \cite{BlumeKohout2006Accurate}) the unique Bayes estimator is the mean of the posterior distribution
\[
\hat{\vec{x}}(D;C) = \mathbb{E}_{\vec{x}|D;C}[\vec{x}].
\]
Minimizing the Bayes risk of a choice of parameters is equivalent to maximizing the negative Bayes risk for that set; therefore,
it is reasonable to choose the negative Bayes risk as our utility function.  It also has theoretical benefits in that it is easy
to compare the performance of algorithms that take $U(c_{N+1})=-r(\pi;c_{N+1}, C)$.


The question of how well can we estimator $\vec{x}$ becomes the question of how low can we make the Bayes risk $r(\pi;C)$.   We lower bound the achievable risk via the Bayesian variant of the Cramer-Rao bound \cite{Gill1995Applications}.   Both require finding the Fisher information. In the case of multiple parameters, the Fisher information is a matrix defined by
\[
  \matr{I}(\vec{x};C) = \mathbb{E}_{D|\vec{x};C} \left[
    \vec{\nabla}_{\vec{x}}      \log\left(\Pr(D|\vec{x};C)\right) \cdot
    \vec{\nabla}_{\vec{x}}^{\T} \log\left(\Pr(D|\vec{x};C)\right)
  \right].
\]
The Fisher information does not depend at all on the prior distribution, and thus is calculated in the same way regardless of how many experiments have already been performed.

The standard Cramer-Rao bound is then given by $\Cov(\vec{\hat{x}}) \ge \matr{I}(\vec{x};C)^{-1}$, where $\matr{X} \ge \matr{Y}$ means that $\matr{X} - \matr{Y}$ is positive semi-definite.  If we choose  the matrix $\matr{Q}$ associated with the quadratic loss to be $\matr{Q}= \id$, then $R(\vec x, \vec{\hat{x}};C) = \Tr(\Cov(\hat{\vec{x}})) \ge \Tr(\matr{I}(\vec{x};C)^{-1})$.  Clearly, this statement of the multivariate Cramer-Rao bound assumes that $\matr{I}$ is non-singular. 
Singular Fisher information matrices arise when there are experiments that provide no information about \emph{at least one} of the experimental parameters.
Unfortunately, that assumption is not met in general.  We avoid this problem by considering the Bayesian information matrix $\matr{J}(\pi;C) = \mathbb{E}_{\vec{x}}[\matr{I}(\vec{x};C)] $.  Then, the \emph{Bayesian Cramer-Rao bound} (BCRB) is given by \cite{Gill1995Applications}
\[
r(\pi;C) \ge \matr{J}(\pi;C)^{-1}.
\]
Lower bounds can be found for specific values of $C$ using numerical integration.  Here we apply an iterative algorithm\footnote{This type of algorithm has been given some attention recently for ``state space models'' and classical signals with additive noise \cite{Tichavsky1998Posterior}.} to sequentially monitor the lower bound.  We will subscript the various quantities of interest by $N$, which means ``at the $N^{\text{th}}$ measurement''.  Then, the Bayesian Cramer-Rao bound is given by
\begin{equation}\label{BCRB-iterate}
r(\pi;c_{N+1})\geq \matr{J}_{N+1}(c_{N+1})^{-1},
\end{equation}
where the iteration is given by
\[
\matr{J}_{N+1}(c_{N+1}) = \matr{J}(\pi;c_{N+1}) + \matr{J}_{N}(c_{N})
\]
and the initial condition is $\matr{J}_0 = \mathbb{E}_{\vec{x}}[\vec{\nabla}_{\vec{x}}      \log\left(\pi(\vec{x})\right) \cdot
    \vec{\nabla}_{\vec{x}}^{\T} \log\left(\pi(\vec{x})\right)]$.

\section{Sequential Monte Carlo Algorithm}
\label{sec:SMC}

Monte Carlo methods remedy problems that deterministic numerical integrators encounter in finding the volume of multi--dimensional spaces.  Specifically, the errors incurred by Monte Carlo integrators do not depend on the dimension of the space.  The variant of
Monte Carlo integration that we use dates back to 1993 \cite{Gordon1993Novel} but has been rediscovered many times in a wide variety of scientific applications under the following names\footnote{See, for example, \cite{Doucet2009Tutorial} for a recent tutorial on these methods.}: sequential Monte Carlo, particle filters, sequential importance sampling, Bayes filters, and so on. We will use the term sequential Monte Carlo (SMC) and will now sketch the idea behind the algorithm.

Recall the Bayes update rule for one datum $d_1$,
\[
\Pr(\vec{x}|d_1) \propto \Pr(d_1|\vec{x}) \Pr(\vec{x}).
\]
The distribution can be processed sequentially as more data arrive:
\begin{align*}
\Pr(\vec{x}|d_2,d_1) &\propto \Pr(d_2|\vec{x}) \Pr(\vec{x}|d_1),\\
\Pr(\vec{x}|d_3,d_2,d_1) &\propto \Pr(d_3|\vec{x}) \Pr(\vec{x}|d_2,d_1),\\
&\vdots
\end{align*}
These updates are difficult to perform because the evaluation of the posterior distributions require evaluations
of the costly multi--dimensional integrals over the parameter space.
We address this issue by using \emph{sequential Monte Carlo} methods, which approximate a distribution over parameters with a distribution that has support only over a finite number of points (often referred to as \emph{particles}). The support of
the distribution over these points is called the weight of the \emph{particle}, and by convention the sum of all weights must
be $1$. More concretely, we approximate an arbitrary distribution by
\[
\Pr(\vec{x}|D) \approx \sum_{k=1}^n w_k(D) \delta(\vec{x} - \vec{x}_k),
\]
where the weights at each step are iteratively calculated from the previous step via
\[
w_k(d_{j+1}\cup D) = \sum_{k=1}^n \Pr(d_{j+1}|\vec{x}_k)w_k(d_j).
\]
 Using this form of a SMC algorithm
also has the advantage of ensuring that the prior and posterior distributions for the update always have support over a finite
number of particles, which simplifies the analysis of our algorithm.

The particle approximation can be made arbitrarily accurate by increasing the number of particles and
 will be a good approximation at every update provided we feed in, at the initial stage, the appropriate weights $\{w_k\}$ and support points  $\{\vec{x}_k\}$.
Since both the weights and support points of the particles carry information about distributions over the model parameters $\vec{x}$, we can without loss of generality
choose the initial weights to be uniform, $w_k = 1/n$ for all $k$, and the initial support points to be samples from the correct prior $\Pr(\vec{x})$. Having made the particle approximation, we can compute expectation values and variances according to Algorithms \ref{alg:smc-est} and \ref{alg:smc-meanfn}, and can perform Bayes updates according to Algorithm \ref{alg:smc-update}. 

\begin{algorithm}
\caption{Estimators for mean and covariance using Sequential Monte Carlo.}
\label{alg:smc-est}
\begin{algorithmic}
\Require Particle weights $w_i$, $i \in \{1, \dots, n\}$.
\Require Particle locations $\vec{x}_i$,  $i \in \{1, \dots, n\}$.
\Ensure Approximations $\vec{\mu}$ and $\matr{\Sigma}$ of $\expect[\vec{x}]$ and $\Cov(\vec{x})$.
\Function {Mean}{$\{w_i\}$, $\{\vec{x}_i\}$}
  \State \Return $\vec{\mu} \gets \sum_i w_i \vec{x}_i$
\EndFunction
\\
\Function {Cov}{$\{w_i\}$, $\{\vec{x}_i\}$}
\Comment Estimates the covariance matrix $\Cov(\vec{x}) = \expect[\vec{x} \vec{x}^{\T}] - \expect[\vec{x}] \expect[\vec{x}]^{\T}$.
  \State $\vec{\mu} \gets \sum_i w_i \vec{x}_i$
  \State $\matr{\Sigma} \gets \sum_i w_i \vec{x}_i \vec{x}_i^{\T} - \mu \mu^{\T}$
  \State \Return $\matr{\Sigma}$
\EndFunction
\end{algorithmic}
\end{algorithm}

\begin{algorithm}
\caption{Estimator for arbitrary expectation values using SMC.}
\label{alg:smc-meanfn}
\begin{algorithmic}
\Require Particle weights $w_i$, $i \in \{1, \dots, n\}$.
\Require Particle locations $\vec{x}_i$,  $i \in \{1, \dots, n\}$.
\Require Function $f(\vec{x})$ of model parameters to average over.
\Ensure Approximation $\expect[f(\vec{x})]$.

\Function {MeanFn}{$\{w_i\}$, $\{\vec{x}_i\}$, $f$}
  \State \Return $\sum_i w_i f(\vec{x}_i)$
\EndFunction

\end{algorithmic}
\end{algorithm}

\begin{algorithm}
\caption{Sequential Monte Carlo update algorithm.}
\label{alg:smc-update}
\begin{algorithmic}
\Require Particle weights $w_i$, $i \in \{1, \dots, n\}$.
\Require Particle locations $\vec{x}_i$,  $i \in \{1, \dots, n\}$.
\Require New datum $D$, obtained from an experiment with control $C$.
\Ensure  Updated weights $w_i'$.
\Function{Update}{$\{w_i\}$, $\{\vec{x}_i\}$, $D$, $C$}
  \For{$i \in 1 \to n$}
    \State $\tilde{w}_i \gets w_i \Pr(D | \vec{x}_i, C)$
    \Comment The updated weights $\tilde{w}_i$ are unnormalized.
  \EndFor
  \State $w'_j \gets \tilde{w}_j / \sum_i \tilde{w}_i$
  \Comment We must normalize the updated weights before returning.
  \State \Return $\{w'_j\}$
\EndFunction
\end{algorithmic}
\end{algorithm}

Sequential Monte Carlo techniques require careful effort to avoid introducing errors due to limited numerical precision. The first problem any SMC algorithm runs into is zero weights.  This is doubly painful since we are effectively operating with fewer particles but using the same amount of computational resources. Since the support of our approximate distribution is a measure-zero set according to the correct distribution, all the weights will eventually be zero; we cannot avoid this but
it can be postponed by using \emph{resampling} techniques.

Generally, the idea behind resampling is to adaptively change the location of the particles to those which are most likely.  The simplest of these types of algorithm chooses $n$ particles (the original number), with replacement, according to the distribution of weights then reset the weights of all particles to $1/n$.  Thus, zero weight particles are ``moved'' to higher weight locations. To determine when to resample, we shall compare the effective sample size $n_{\text{ess}} = 1/\sum_i w_i^2$ to a threshold \verb"resample_threshold", which is the effective ratio of the original number of particles $n$. We use \verb"resample_threshold"$ = 0.5$, as suggested by~\cite{Liu2000Combined}.

The resampling algorithm we use was first proposed in~\cite{Liu2000Combined} and is given explicitly in Algorithm~\ref{alg:smc-resample}.  The idea behind the algorithm conforms to
the intuition given above but it incorporates randomness to search larger volumes of the parameter space.  This randomness is inserted in the resampling algorithm by
applying a random perturbation to the location of each particle that is introduced during the resampling process.  Thus, the new particles are randomly spread around the previous locations of the old.  More formally, we model this by randomly choosing a particle location $\vec{x}_i$, then perturbing it by a normally distributed vector $\vec{\epsilon}\sim\mathcal{N}(0, \Sigma)$ (we will come back to how to choose the mean and covariance). The new particles are thus samples of the convolved distribution
\begin{equation}
	p(\vec{x}') = \sum_i w_i \frac{1}{\sqrt{(2\pi)^k\vert\Sigma\vert}} \exp\left(
	  -\frac{1}{2} {(\vec{x}'-\vec{\mu}_i)}^{\T} \Sigma^{-1} (\vec{x}'-\vec{\mu}_i)
	\right),\label{eq:sec4:seqprob}
\end{equation}
where $k$ is the number of model parameters. A distribution of this form is known as a \emph{mixture distribution}, and can be efficiently sampled by first choosing a particle, then choosing a perturbation vector.

To choose the mean $\vec{\mu}_i$ of each term in the resampling mixture distribution, we choose a vector that is a convex combination of the original particle location $\vec{x}_i$ and the expected model $\vec{\mu} = \expect[\vec{x}]$, so that
\[
	\vec{\mu}_i = a \vec{x}_i + (1 - a) \vec{\mu},
\]
where $a$ is a tunable parameter of the resampling algorithm. We will use $a = 0.98$, as suggested by \cite{Liu2000Combined}. The covariance of each perturbation is then given by
\[
	\matr{\Sigma} = (1 - a^2) \Cov[\vec{x}].
\]
Our resampling algorithm then involves drawing $n$ new particles from the distribution given by \eqref{eq:sec4:seqprob} and setting the weight of each new particle to $1/n$.

\begin{algorithm}
\caption{Sequential Monte Carlo resampling algorithm.}
\label{alg:smc-resample}
\begin{algorithmic}
\Require Particle weights $w_i$, $i \in \{1, \dots, n\}$.
\Require Particle locations $\vec{x}_i$,  $i \in \{1, \dots, n\}$.
\Require Resampling parameter $a \in [0, 1]$.
\Ensure  Updated weights $w_i'$ and locations $\vec{x}'_i$.
\Function{Resample}{$\{w_i\}$, $\{\vec{x}_i\}$, $a$}
  \State $\vec{\mu} \gets$ \Call{Mean}{$\{w_i\}, \{\vec{x}_i\}$}
  \State $h \gets \sqrt{1-a^2}$
  \State $\matr{\Sigma} \gets h^2$ \Call{Cov}{$\{w_i\}, \{\vec{x}_i\}$}
  \For{$i \in 1 \to n$}
    \State draw $j$ with probability $w_j$
    \Comment Choose a particle $j$ to perturb.
    \State $\vec{\mu}_i \gets a\vec{x}_j + (1-a) \vec{\mu}$
    \Comment Find the mean for the new particle location.
    \State draw $\vec{x}'_i$ from $\mathcal{N}(\vec{\mu_i}, \matr{\Sigma})$
    \Comment Draw a perturbed particle location.
    \State $w_i \gets 1/n$
    \Comment Reset the weights to uniform.
  \EndFor

  \State \Return $\{w'_i\}$, $\{\vec{x}'_i\}$
\EndFunction
\end{algorithmic}
\end{algorithm}


There are a few details to address regarding the efficiency of the SMC algorithm.  The first thing to note is that, since the choice of resampling algorithm is usually tailored to the problem at hand, it is hard to say something in general about the algorithmic complexity of it.  A more pressing issue for us, however, is that quantum simulation is required in order to evaluate
the likelihood function.  This step will not generally be efficient because no known classical algorithm exists that can simulate
generic quantum dynamics in time polynomial in the number of interacting subsystems.

Thus, since calls to the likelihood function are expensive, we wish to minimize the number of times it is called.  To achieve this, we use an approximation that involves using only the highest weighted particles to compute the expectation values appearing in the utility function.  Note that this will not reduce the accuracy of the estimation of parameters \emph{given} experiments -- rather, it reduces the accuracy with which we choose optimal experiments.  This is the ideal place to make the approximation since the optimization routine makes the most calls to the likelihood function and no experiment is ``bad'' in the sense that the expected risk (and hence the actual risk, on average) can increase.  This idea is formalized in the algorithm by setting a parameter \verb"approx_ratio" which is the percentage of particles to use for the approximation.  We give pseudocode for this algorithm
in Algorithm~\ref{alg:smc-approxutils}.

\begin{algorithm}
\caption{Approximate utility functions using Sequential Monte Carlo.}
\label{alg:smc-utility}
\begin{algorithmic}
\Require Particle weights $w_i$, $i \in \{1, \dots, n\}$.
\Require Particle locations $\vec{x}_i$,  $i \in \{1, \dots, n\}$.
\Require Control description $C$.
\Require Positive semi-definite scaling matrix $\matr{Q}$.
\Ensure  Utility $U(C)$.

\Function{UtilNV}{$\{w_i\}$, $\{\vec{x}_i\}$, $C$, $\matr{Q}$}
\Comment Calculates the negative variance utility function.

  \For{$D \in 1 \to n_{\text{outcomes}}$}
    \State $\{w'_i\} \gets$ \Call{Update}{$\{w_i\}$, $\{\vec{x}_i\}$, $D$, $C$}
    \Comment Find the hypothetical weights, had we obtained the datum $D$.
    \State $\vec{\mu} \gets$ \Call{Mean}{$\{w'_i\}$, $\{\vec{x}_i\}$}
    \Comment Calculate the mean using the updated weights.
    \State $u_D \gets -\sum_i w_i (\vec{x}_i - \vec{\mu})^{\T} \matr{Q} (\vec{x}_i - \vec{\mu})$
    \Comment Reduce the variance to a scalar by using the scaling matrix $\matr{Q}$.
    \State $u_D \gets u_D \cdot \sum_i w_i \Pr(D | \vec{x}_i, C)$
    \Comment Weight the partial utility $u_D$ by the marginalized likelihood $\Pr(D | C)$.
  \EndFor

  \State \Return $\sum_D u_D$
\EndFunction
\\
\Function{UtilIG}{$\{w_i\}$, $\{\vec{x}_i\}$, $C$}
\Comment Calculates the information gain utility function.

  \For{$D \in 1 \to n_{\text{outcomes}}$}
    \For{$i \in 1\to n$}
      \State $p_{D|\vec{x}_i} \gets \Pr(D | \vec{x}_i, C)$
    \EndFor
    \State $p_D \gets \sum_i w_i p_{D|\vec{x}_i}$
  \EndFor

  \State \Return $I \gets H_D(p_D) - \sum_i w_i H_D(p_{D | \vec{x}_i})$
  \Comment $H_D$ is the entropy over data $D$.

\EndFunction
\end{algorithmic}
\end{algorithm}

\begin{algorithm}
\caption{Reduced particle approximation for Sequential Monte Carlo utility functions.}
\label{alg:smc-approxutils}
\begin{algorithmic}
\Require Particle weights $w_i$, $i \in \{1, \dots, n\}$.
\Require Particle locations $\vec{x}_i$,  $i \in \{1, \dots, n\}$.
\Require Ratio \texttt{approx\_ratio} of the particles to keep in the reduced approximation.

\Ensure  Reduced sets of particle weights $\{\tilde{w}_i\}$ and locations $\{\tilde{\vec{x}}_i\}$.

\Function{Reapprox}{$\{w_i\}$, $\{\vec{x}_i\}$, \texttt{approx\_ratio}}

  \State $\tilde{n} \gets \left\lfloor n\cdot \text{\texttt{approx\_ratio}} \right\rfloor$
  \State draw $\pi$ uniformly at random from $\Sym(n)$
  \Comment $\Sym(n)$ is the symmetric group acting on $n$ elements.
  \State $\{\tilde{w}_i\} \gets \{w_{\pi(i)}\}$
  \Comment We permute the elements to avoid introducing patterns when sorting the particle weights.
  \State $\{\tilde{\vec{x}}_i\} \gets \{\vec{x}_{\pi(i)}\}$
  \State $\{s_k\} \gets $ \Call{Sort}{$\{\tilde{w}_{i}\}$}
  \Comment Get a list of indices $s_i$ such that $\tilde{w}_{s_i} \ge \tilde{w}_{s_j}$ for all $i$, $j$.

  \State \Return $\{\tilde{w}_i\} \gets \{\tilde{w}_{s_i} : i \in 1 \to \tilde{n}\}$,
      $\{\tilde{\vec{x}}_i\} \gets \{\tilde{\vec{x}}_{s_i} : i \in 1 \to \tilde{n} \}$

\EndFunction
\end{algorithmic}
\end{algorithm}

We combine these prior algorithms to obtain Algorithm \ref{alg:smc-complete}, which is our complete algorithm for adaptively designing experiments using the SMC approximation. Note that we have left unspecified here the choice of local optimizer; in practice, this will be chosen depending on what works for a given experimental model. In Section \ref{sec:benchmarking}, we compare the Newton conjugate-gradient (NCG) and nonlinear conjugate-gradient methods to a ``null'' optimizer that only evaluates the utility function at the initial guesses. Also in Section \ref{sec:benchmarking}, we consider which choices of $n$, $n_{\text{guesses}}$ and \verb+approx_ratio+ result in a useful estimation algorithm.

\begin{algorithm}
\caption{Complete adaptive Bayesian experiment design algorithm, using sequential Monte Carlo approximations.}
\label{alg:smc-complete}
\begin{algorithmic}
\Require A number of particles $n$ to be used.
\Require A prior distribution $\pi$ over models.
\Require A number of experiments $N$ to perform.
\Require A resampling parameter $a\in[0,1]$.
\Require A threshold \verb"resample_threshold" $\in[0, 1]$ specifying how often to resample.
\Require An approximation ratio \verb"approx_ratio".
\Require An local optimization algorithm \textsc{LocalOptimize}.
\Require A particular choice of utility function \textsc{Util}.
\Require A heuristic \textsc{GuessExperiment} for choosing experiment controls, and a number $n_{\text{guesses}}$ of potential experiments to consider in each iteration.
\Ensure  An estimate $\hat{\vec{x}}$ of the true model $\vec{x}_0$.
\Statex 
\Function{EstimateAdaptive}{$n$, $\pi$, $N$, $a$, \texttt{resample\_threshold}, \texttt{approx\_ratio}, \textsc{Optimize}, \textsc{Util}, $n_{\text{guesses}}$, \textsc{GuessExperiment}}
  \Statex
  \State  $w_i \gets 1/n$
  \Comment Start by initializing the SMC variables.
  \State  draw each $\vec{x}_i$ independently from $\pi$

  \Statex
  \For {$i_{\text{exp}} \in 1\to N$}
  \Comment We now iterate through each experiment.
    \If {\texttt{approx\_ratio} $\ne 1$}
    \Comment If we are using a reduced particle set, populate that first.
      \State $\{\tilde{w}_i\}$, $\{\tilde{\vec{x}}_i\}\gets$ \Call{Reapprox}{$\{w_i\}$, $\{\vec{x}_i\}$, \texttt{approx\_ratio}}
    \Else
      \State $\{\tilde{w}_i\}$, $\{\tilde{\vec{x}}_i\}$ $\gets$ $\{w_i\}$, $\{\vec{x}_i\}$
    \EndIf

    \Statex
    \For {$i_{\text{guess}} \in 1 \to n_{\text{guesses}}$}
      \Comment Heuristicly choose potential experiments, and optimize each independently.
      \State $C_{i_{\text{guess}}} \gets $ \Call{GuessExperiment}{$i_{\text{exp}}$}
      \State $\hat{C}_{i_{\text{guess}}}$, $U_{i_{\text{guess}}}$ $\gets$ \Call{LocalOptimize}{\textsc{Util}, $C_{i_{\text{guess}}}$, $\{\tilde{w}_i\}$, $\{\tilde{\vec{x}}_i\}$}
    \EndFor

    \Statex
    \State  $i_{\text{best}} \gets \operatorname{argmax}_{i_{\text{guess}}} U_{i_{\text{guess}}}$
    \Comment We pick the experiment whose post-optimization utility is highest.
    \State  $\hat{C} \gets \hat{C}_{i_{\text{best}}}$

    \State  $D_{i_{\text{exp}}} \gets $ the result of performing $\hat{C}$
    \Comment The best experiment is then performed.
    \State  $\{w_i\}$, $\{\vec{x}_i\}$ $\gets$ \Call{Update}{$\{w_i\}$, $\{\vec{x}_i\}$, $D$, $C$}
    \Comment This update carries the posterior distribution forward.

    \Statex
    \If {$\sum_i w_i^2 < N\cdot\text{\texttt{resample\_threshold}}$}
    \Comment Resample if the effective sample size $n_{\text{ess}} <$ \verb"resample_threshold".
      \State  $\{w_i\}$, $\{\vec{x}_i\}$ $\gets$ \Call{Resample}{$\{w_i\}$, $\{\vec{x}_i\}$, $a$}
    \EndIf
  \EndFor

  \Statex
  \State \Return $\hat{\vec{x}} \gets $\Call{Mean}{$\{w_i\}$, $\{\vec{x}_i\}$}
  \Comment After all experiments have been performed, return the mean as an estimate.
  \Statex

\EndFunction
\end{algorithmic}
\end{algorithm}

\section{Region Estimation}
\label{sec:region-est}

In addition to providing an accurate estimate of the true model parameters for the system, it is important to be able to quantify
the uncertainty in the estimated model parameters.  This task can be achieved by finding a region $\hat{X}$
of the space of models such that $\Pr(\vec{x}_0\in\hat{X})$ is maximized and such that $\Vol(\hat{X})$ is minimized. This is useful, for instance, if
we consider the region estimate $\hat{X}$ as input to an optimal control theory (OCT) algorithm that describes the range of dynamics experienced by
a quantum system. Since the OCT algorithm must find a control design that is robust for every point in that range, the cost of that optimization
increases with the volume of $\hat{X}$.

Because we are interested in the probability $\Pr(\vec{x}_0 \in \hat{X})$ of the true model $\vec{x}_0$ lying within our region estimate $\hat{X}$
for a given run of the SMC algorithm, we say that $\hat{X}$ is a \emph{credible} region \cite{edwards_bayesian_1963, bernardo_intrinsic_2005}.
This is in contrast to \emph{confidence} regions which are functions from data records $D$ to regions
$\hat{X}_{\textrm{conf}}(D)$ such that the probability of obtaining a data record
for which $\vec{x}_0 \in \hat{X}_{\textrm{conf}}(D)$ is at least some threshold \cite{wackerly_mathematical_2001, beale_confidence_1960}.
That is, credible regions give probabilities about a single data record,
while confidence regions give probabilities about all possible data.
Broadly speaking, credible regions are Bayesian analogues to the frequentist concept of confidence regions.
Unless otherwise noted, we concern ourselves here with credible regions as obtained from posterior distributions.
The two kinds of region estimates are closely related, as has recently
been explored in the context of quantum tomography \cite{BlumeKohout2012Robust,Christandl2011Reliable}.

We make the problem of finding a credible region estimate amenable to analysis by SMC by
turning the problem into that of estimating an expectation value.
In particular, the probability of the true model being within a region can be expressed as
\[
  \Pr(\vec{x}_0 \in \hat{X}) = \expect[1_{\hat{X}}],
\]
where $1_{\hat{X}}$ is the \emph{indicator function} for $\hat{X}$, defined by
\[
  1_{\hat{X}} (\vec{x}) = \begin{cases}
    1 & \vec{x} \in \hat{X} \\
    0 & \vec{x} \notin \hat{X}
  \end{cases}.
\]
The expectation value $\expect[1_{\hat{X}}]$ can then be computed using Algorithm~\ref{alg:smc-meanfn}, giving that
\begin{align*}
  \expect[1_{\hat{X}}] & \approx \sum_i w_i 1_{\hat{X}} (\vec{x}_i) \\
                       & =       \sum_{i, \vec{x}_i \in \hat{X}} w_i.
\end{align*}

Thus, by construction, any region containing particles of total weight at least $r$ will have an approximate probability mass of at least $r$.
We formalize this intuition by introducing a \emph{probability mass} function $m(R)$ on regions $R$ such that
\[
  m(R) = \expect[1_R].
\]
Similarly, let $\tilde{m}(R) = \sum_{i,\ i\in R} w_i$ be an approximation of $m(R)$ using the SMC algorithm.

We thus seek a region $\hat{X}$ such that $\Vol(\hat{X})$ is small, $m(\hat{X})$ is large and such that $\hat{X}$ is an efficiently computable property
of the current SMC state. We achieve the latter two properties by choosing some appropriate geometric function of
a set of particles $X_r$ whose weight is above some threshold weight $r$; for example, the convex hull or the minimum-volume enclosing ellipse of $X_r$
both satisfy $\tilde{m}(X_r) \ge r$ and may be computed using well-known classical algorithms \cite{todd_khachiyans_2007,barber_quickhull_1996}.


We improve on these results by supposing that, after collecting a reasonable amount of data, the posterior distribution is approximately normally distributed according to $\mathcal{N}(\vec{\mu}, \matr{\Sigma})$ for some $\vec{\mu}$ and $\matr{\Sigma}$.  This assumption holds when the Fisher information is
non--singular, and we find in the case of our benchmarks that it approximately holds if $\Pr(\vec{x}|D)$ is sharply peaked.
Under this assumption, it follows from the definition of the multivariate normal distribution that the inverse covariance matrix $\matr{\Sigma}^{-1}$ describes an ellipse such
that approximately $0.682^d$ of the probability mass is contained inside the ellipse, where $d$ is the number of unknown model parameters, $d = \dim \vec{x}$. In particular, the covariance matrix
transforms a vector $\vec{z}$ of length $d$ with each component drawn from the standard normal distribution $\mathcal{N}(0, 1)$ into a random variate of a
multivariate normal distribution with the given covariance, and so inverting that transformation gives the $z$-score for each component.

More generally, under the assumption of a normally distributed posterior, the error ellipse of points $\vec{x}$ satisfying
\begin{equation}
  (\vec{x}-\vec{\mu})^{\T} \matr{\Sigma}^{-1} (\vec{x}-\vec{\mu}) \le Z^2\label{eq:ellipse}
\end{equation}
for some $Z>0$ will contain a ratio
\[
  \left(\cdf_{\mathcal{N}}(Z) - \cdf_{\mathcal{N}}(-Z)\right)^d = \erf\left[\frac{Z}{\sqrt{2}}\right]^d
\]
of the particle weight, where $\cdf_{\mathcal{N}}(Z)$ is the cumulative distribution function for the normal distribution, evaluated at $Z$. Thus, if the assumption of a normal posterior is a good approximation, then
the estimated covariance matrix according to Algorithm \ref{alg:smc-est} can be used as a region estimator.

The volume of the covariance ellipse region estimator can then be found by again treating $\matr{\Sigma}$ as a transformation of a $d$-dimensional coordinate system, so that
\[
  \Vol(\matr{\Sigma}^{-1}) = \frac{\pi^{d/2}}{\Gamma(\frac{d}{2} + 1)} \det(\matr{\Sigma}^{1/2}).
\]

We test that the normal posterior assumption is approximately met by finding the approximate probability mass $\tilde{m}(\Cov(\hat{\vec{x}})^{-1}/Z^2)$ for a given $Z$,
and comparing to the expected cdf. Moreover, we compare the optimal size of the covariance ellipse to the actual size by appealing to the Bayesian Cramer-Rao bound,
since $\expect_{\pi}[\Cov(\hat{\vec{x}})] \ge \matr{J}(\pi; C)^{-1}$. This comparison will be explored further in Section \ref{sec:benchmarking}, and will be used as a
performance metric for our algorithm.


\section{Hyperparameter Estimation}
\label{sec:hyperparam-est}
Now that we have discussed the general framework for using Bayesian inference to learn Hamiltonian
parameters, we will proceed to discuss an important generalization of the prior work.  The generalization
that we consider addresses the fact that quantum systems seldom have consistent Hamiltonians from
experiment to experiment, due to experimental errors.  
Hyperparameters allow us to generalize the Hamiltonian learning problem from one involving learning
the Hamiltonian parameters to one that involves learning the parameters that describe the distribution 
of Hamiltonian parameters. In this way, the method of hyperparameter estimation is an alternative
to approaches such as quantum smoothing \cite{Tsang2009TimeSymmetric,Wheatley2010Adaptive}, which
integrate over the \emph{history} of a time-varying random parameter, rather than considering 
the distribution from which each realization is being drawn.

We denote the hyperparameters for a model Hamiltonian as $\vec{y}$ to avoid subtle conceptual
differences between the hyperparameters and the distributions on $\vec{x}$ that they describe.
The probability distribution for $\vec{x}$ can then be written as $\Pr(\vec{x}|\vec{y})$.
Despite interpretational differences, the hyperparameters can also be learned using Algorithm~\ref{alg:smc-complete}
in exactly the same way that $\vec{x}$ is learned.  The region estimates yielded by the algorithm are 
region estimations for $\vec{y}$ and, as we will show shortly, can easily be converted into region estimates
for $\vec{x}$.

The drawback to this approach is that computations of the likelihood function can become much more
expensive.  Specifically, in order to compute the likelihood function $\Pr(D|\vec{y})$ we need to compute the probabilities
of data $d$ emerging from an ensemble of randomly sampled experimental parameters taken from
the distribution described by $\vec{y}$.  A large number of samples, $N_s$, may be required in some cases
since the sample error scales as $1/\sqrt{N_s}$.  On the other hand, the approach is straight forward
to implement because it does not require either increases to the number of particles or changes to the
underlying algorithm.

In some important special cases, this drawback can be avoided by analytically performing the marginalization over $\vec{x}$,
\[
  \Pr(D|\vec{y}) = \int \dif\vec{x} \Pr(D|\vec{x}) \Pr(\vec{x} | \vec{y}).
\]
In Section \ref{sec:hyperparameter-model}, we discuss a particular case where the marginalization is analytically tractable.

The resulting means and covariance matrices for $\vec{y}$ can be readily converted to the corresponding
quantities for $\vec{x}$ by using the chain rule for expectation values,
\begin{equation}
\expect_{\vec{x},\vec{y}}[\vec{x}]= \expect_{\vec{y}}[\expect_{\vec{x}|\vec{y}}[\vec{x}]].
\end{equation}
This expectation value can be computed using the posterior distribution $\Pr(\vec{y}|D)$ and the intermediate model distribution $\Pr(\vec{x}|\vec{y})$, which will typically be easy to compute from the definition of the hyperparameters.
The covariance matrix for $\vec{x}$ is slightly more complicated.  It is straightforward to verify that
\begin{equation}
\Cov_{\vec{x}, \vec{y}} (\vec{x}) = \expect_{\vec{y}}\left[\Cov_{\vec{x}|\vec{y}}(\vec{x})\right] +\Cov_{\vec{y}}\left(\expect_{\vec{x}|\vec{y}}[\vec{x}]\right).
\label{eq:hyperregion-cov}
\end{equation}
For the special case that $\vec{x}$ is a single parameter, the covariance can be replaced with the variance to obtain that
\begin{equation}
\Var_{\vec{x}, \vec{y}} (\vec{x}) = \expect_{\vec{y}}\left[\Var_{\vec{x}|\vec{y}}(\vec{x})\right] +\Var_{\vec{y}}\left(\expect_{\vec{x}|\vec{y}}[\vec{x}]\right).
\label{eq:hyperregion-var} 
\end{equation}
Using the region estimate given by~\eqref{eq:ellipse} to estimate a hyperparameter region translates to a region estimator for the model parameters $\vec{x}$, if the distribution over hyperparameters $\vec{y}$ is approximately Gaussian near its peak. In the limit of many experiments, we find that this is a good assumption, as is discussed in Section \ref{sec:benchmarking}.

An important consequence of this derivation is that the same SMC algorithm can be used to estimate regions of model parameters, even when those model parameters vary with each experiment. In particular, as the covariance $\Cov_{\vec{y}}(\hat{\vec{x}})$ in our estimate $\hat{\vec{x}}$ of the mean model parameter vector decreases, our estimate of the model region approaches the ``true'' model region given by $\Cov_{\vec{x}|\vec{y}}(\vec{x}|\vec{y})$.

\section{Test Cases}
\label{sec:test cases}
We assess the performance of Algorithm~\ref{alg:smc-complete} through a number of test cases
that are designed to examine the performance of the algorithm in a number of different relevant
settings.  Here we describe the test cases, which we build up in complexity starting from the simple model studied in detail in references \cite{Sergeevich2011Characterization,ferrie_adaptive_2012,ferrie_how_2012}.
The first case that we consider is learning a single parameter for an experiment in the presence of a known decoherence time.  The second case generalizes this
by allowing $T_2$ to be unknown.  This problem is particularly important because existing 
experiments require substantial processing to learn both the unknown parameter and the
decoherence time.  Finally, we consider a two--hyperparameter
model with a single-parameter Hamiltonian and perform region estimation of both the hyperparameters and the parameters that
are distributed according to them.

\subsection{Single Parameter Hamiltonian}
\label{sec:single param}
We first consider the example with one unknown and one control parameter.  Suppose that a qubit evolves under an internal Hamiltonian
\[
H(\omega) = \frac{\omega}2 \sigma_z.
\]
Here $\omega$ is an unknown parameter whose value we want to estimate.  An experiment consists of preparing a single known input state $\psi_{\text{in}}=\ket{+}$, the $+1$ eigenstate of $\sigma_x$, evolving under the Hamiltonian $H$ for a controllable time $t$ and performing a measurement in the $\sigma_x$ basis.
The $k^{\text{th}}$ measurement has two outcomes which we record as $d_k\in\{0,1\}$.  The design specification for each experiment
is given by the time $t_k$ that the state is allowed to evolved for under the unknown Hamiltonian in this case.  Protocols that are provably optimal have been obtained by finding analytic expressions, and lower bounds, on the accuracy of generic protocols \cite{ferrie_how_2012}.  

We will slightly generalize this model by allowing noise sources which lead to a decay in the information extractable from any measurement.  This can manifest from, for example, a $T_2$ dephasing process which leads to the following likelihood function:
\begin{align}
\Pr(0|\omega;t) &= e^{-\frac{t}{T_2}}\cos^2\left(\frac\omega 2 t\right)+\frac{1-e^{-\frac{t}{T_2}}}2,\label{SMC1paramModel}\\
\Pr(1|\omega;t) &= 1 - \Pr(0|\omega;t)\nonumber,
\end{align}
where $\omega$ is the unknown parameter to be estimated, $t$ is the controllable parameter and $T_2$ is a known constant.

This model was studied in references \cite{Sergeevich2011Characterization,ferrie_adaptive_2012,ferrie_how_2012}.  There the model was able to be treated analytically.  For the case with no noise ($T_2=\infty$), given a normal prior with variance $\sigma^2$, the risk scales exponentially as $r\sim \sigma^2 (1-e^{-1})^N$, whereas for finite $T_2$, the scaling is exponentially suppressed when the measurement times reach $t=T_2$.


\subsection{Two Parameter Model with Single Control}
\label{sec:two param}
Here, we are going to consider the same model from the previous section,
\begin{align}
\Pr(0|\omega,T_2;t) &= e^{-\frac{t}{T_2}}\cos^2\left(\frac\omega 2 t\right)+\frac{1-e^{-\frac{t}{T_2}}}2,\label{SMC2paramModel}\\
\Pr(1|\omega,T_2;t) &= 1 - \Pr(0|\omega,T_2;t)\nonumber,
\end{align}
but where now both $\omega$ and $T_2$ are unknown.  The time $t$ remains the only experimentally controllable parameter. For numerical convenience, we choose to parameterize this model as $\vec{x}=(\omega, T_2^{-1})$, so that each unknown parameter has the same dimensions.

Even for such a simple generalization as this, the methods discussed above are not adequate for this more general problem.  In particular, the Fisher matrix of any one measurement is singular and hence the standard Cramer-Rao bound does not hold -- nor is it possible to utilize standard asymptotic approximations to normal distributions. Although we cannot find an analytic expression for the Bayesian Cramer-Rao bound in the same way we did for the single parameter problem, we use our SMC algorithm to efficiently numerically estimate it.

In general, there is no reason to expect that $\omega$ and $T_2^{-1}$ will be expressed in such a way as to admit the same scale, and so in estimating using this model, we must set the semidefinite matrix $\matr{Q}$ for the loss function \eqref{eq:mse-loss}. This is discussed further in Section \ref{sec:benchmarking-region-est}.

\subsection{One Parameter Model with Hyperparameters}
\label{sec:hyperparameter-model}

We also derive a two-parameter model similar to \eqref{SMC2paramModel} by again considering a single qubit undergoing Larmor precession as in \eqref{SMC1paramModel}, but where the ``true'' precession frequency $\omega$ is itself distributed according to a Gaussian distribution of mean $\mu$ and variance $\sigma^2$. In this case, following the discussion of Section \ref{sec:hyperparam-est}, the probability of data conditioned on the \emph{hyperparameters} $\vec{y} = (\mu,\ \sigma)$ can be found by marginalizing over the intermediate random variable $\omega$, so that
\begin{align}
  \Pr(d | \mu, \sigma; t) & = \int \Pr(d | \omega) \Pr(\omega | \mu, \sigma) \dif\omega \label{eq:hyperparameter-marginalization-example}. \\
\intertext{
For the specific example of the Gaussian distribution,
}
  \Pr(0 | \mu, \sigma; t) & = \frac{1}{\sigma\sqrt{2\pi}} \int \cos^2\left(\frac{\omega t}{2}\right) e^{-\frac{(\omega - \mu)^2}{\sigma^2}} \dif\omega \label{eq:hyperparameter-model-gaussian} \\
                          & = \frac{1}{2} \left(1 + e^{-2 \sigma ^2 t^2} \cos (2 \mu  t)\right) \label{eq:hyperparameter-model-gaussian-2}.
\end{align}
At this point, we have entirely removed $\omega$ from the problem, leaving a two-parameter model, where we wish to estimate the mean and variance of an unknown normal distribution.

As another example, instead of marginalizing against a Gaussian distribution, we consider the case that the intermediate model parameter $\omega$ is drawn from a Lorentz distribution. A Lorentz distribution is completely determined by its location and scale parameters $\omega_0$ and $\gamma$, respectively, and so we use these hyperparameters to derive a new model,
\begin{align}
  \Pr(0 | \omega_0, \gamma; t) & = \int \cos^2(\omega t / 2) \frac{1}{\pi  \gamma  \left(\frac{\left(\omega -\omega _0\right){}^2}{\gamma ^2}+1\right)} \dif\omega \label{eq:hyperparameter-marginalization-example-cauchy} \\
  & = \frac{1}{2} \left(1 + e^{-t\gamma} \cos \left(t \omega _0\right)\right).
\end{align}
Note that if we identify $\gamma = T_2^{-1}$, then the Lorentz hyperparameter model is the identical to that of Equation \eqref{SMC2paramModel}. This illustrates the relationship between decoherence processes and the lack of knowledge formalized by a hyperparameter model. In a similar fashion, \eqref{eq:hyperparameter-model-gaussian-2} is also model of decoherence. Due to the $t^2$ dependence of the Gaussian-hyperparameter model, \eqref{eq:hyperparameter-model-gaussian-2} represents a decoherence process that  cannot be written in Lindblad form \cite{lindblad_generators_1976} because it cannot be drawn from a quantum dynamical semigroup.

The approach of introducing hyperparameters allows us to estimate unknown distributions using the same SMC algorithm \ref{alg:smc-complete} that we use for estimating other model parameters, by making assumptions about the form of an unknown distribution. Using techniques developed in Section \ref{sec:hyperparam-est}, we can also extend region estimation to hyperparameter models to obtain regions on the intermediate model (in this case, $\omega$) using regions on $\vec{y}$. We discuss the resutls of these application in detail in Section \ref{sec:benchmarking-region-est}.

\section{Results and Discussion}
\label{sec:benchmarking}
We will now turn our attention towards assessing the performance of Algorithm~\ref{alg:smc-complete} 
in practical examples~\cite{imp_details}.
Our results show that our adaptive Bayesian algorithm is able to learn model parameters
using a very small number of experiments, including adverse situations where hyperparameters
are needed to describe the fluctuations of model parameters between experiments or where
the decoherence time of the system is unknown.  This performance is especially noteworthy in 
the case of an unknown decoherence time $T_2$. Though methods of learning unknown decoherence processess have been developed and are well-understood
\cite{Boulant2003Robust,Weinstein2004Quantum,Boulant2004Incoherent},
these methods require multiple iterations of quantum process tomography implemented using either ensemble measurement or a large amount of
data obtained using strong measurement. In the case discussed here, we can adaptively use prior information to obtain
accurate estimates of $T_2$ using significantly less measurements than methods currently employed in systems with strong measurement \cite{SBT11}.

\begin{figure}\centering
  \includegraphics[width=.99\columnwidth]{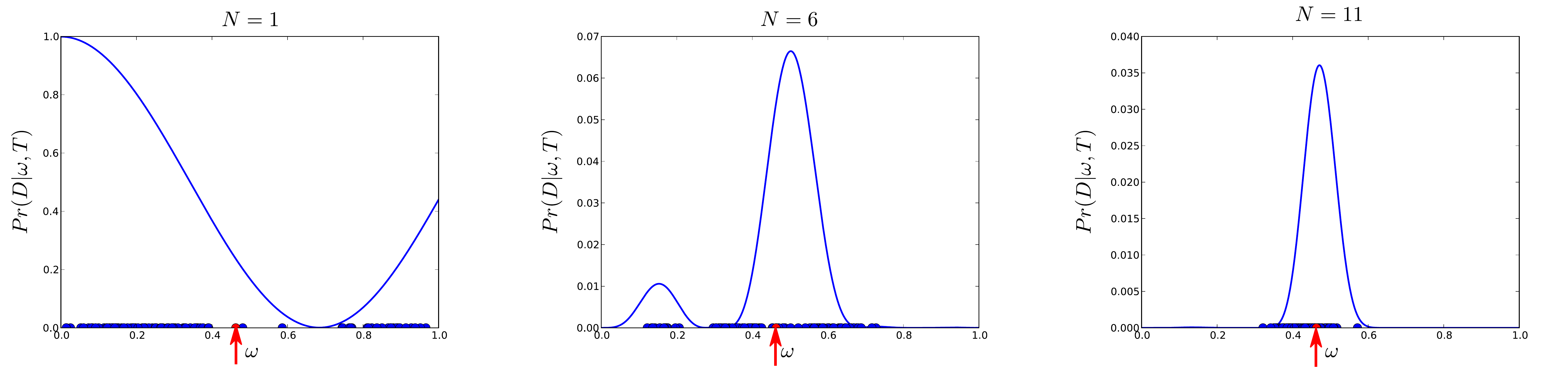}
  \caption{\label{fig:SMC1Dlike} Left to right: the likelihood function for $N = 1, 6,$ and $11$ simulated measurements at random times in in $(0,5\pi)$.  The model is that given in equation \eqref{SMC1paramModel} with $T_2=100\pi$.  The red dots (and red arrows) are the randomly chosen true parameter $\omega$.  The blue dots are the $n = 100$ sequential Monte Carlo ``particles''.}
\end{figure}

\begin{figure}\centering
  \includegraphics[width=.80\columnwidth]{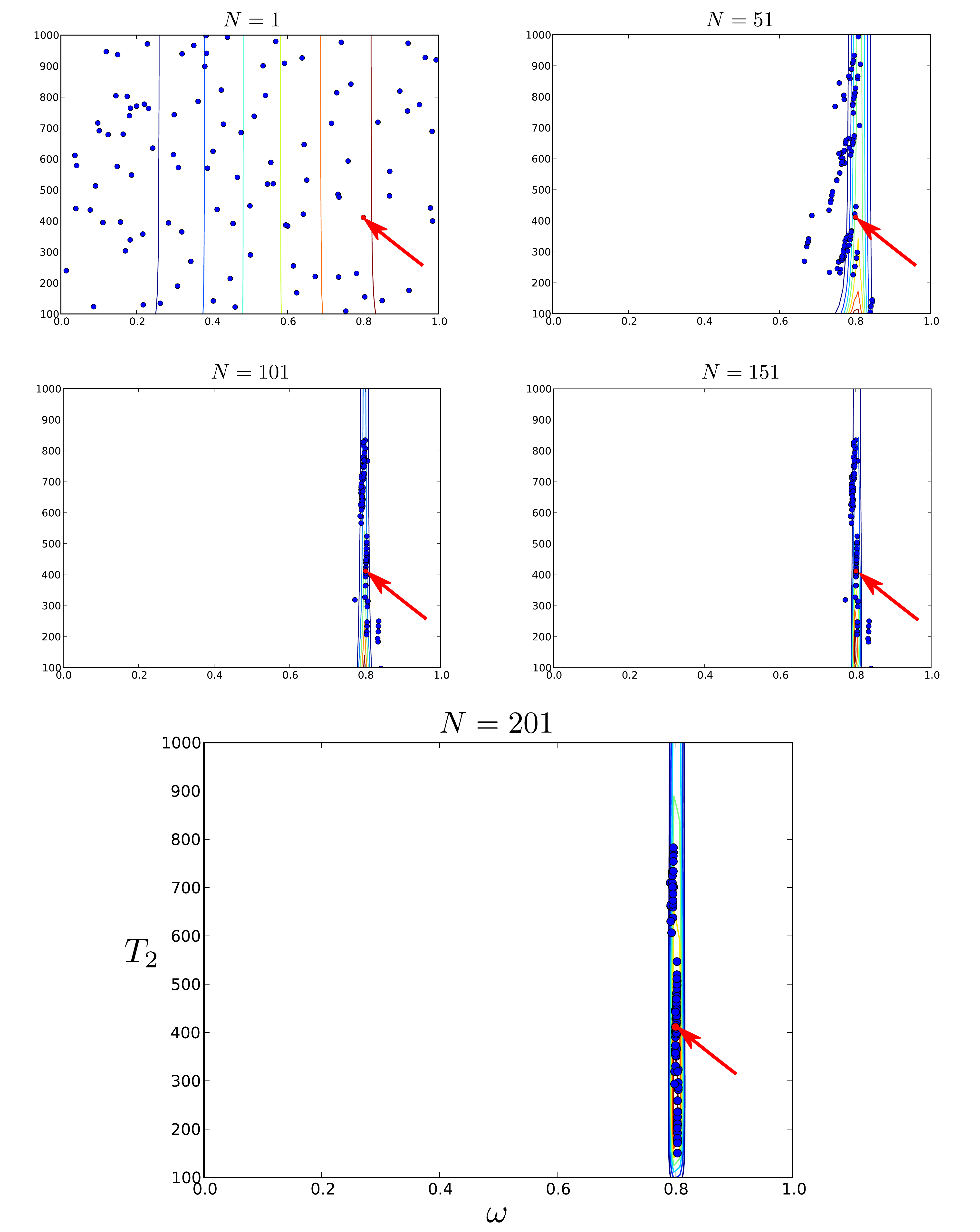}
  \caption{\label{fig:SMC2Dlike} The likelihood function for $N = 1, 51, 101, 151$ and $201$ simulated measurements at random times in in $(0,20\pi)$.  The model is that given in equation \eqref{SMC2paramModel}.  The red dot (and red arrow) is the randomly chosen true parameter $\vec{x} = (\omega,T_2)$.  The blue dots are the $n = 100$ sequential Monte Carlo ``particles''.}
\end{figure}

Figures \ref{fig:SMC1Dlike} and \ref{fig:SMC2Dlike} build intuition for how algorithm \ref{alg:smc-complete} actually learns parameters by describing a trial run for each of the models in Sections \ref{sec:single param} and \ref{sec:two param}.  These illustrate the movement of the particles, through resampling, to regions of high likelihood.  In particular, we note that Fig.~\ref{fig:SMC1Dlike} demonstrates that only $11$
experiments are needed in order to achieve a tight approximately Gaussian posterior distribution over the model parameter $\omega$ for the known $T_2$ model described in Sec.~\ref{sec:single param}.  Figure~\ref{fig:SMC2Dlike} shows
that performance of the algorithm for the unknown $T_2$ model described in Sec.~\ref{sec:two param}.  We see in that case that
only $200$ experiments are needed to find a distribution that is centered around the true values of $\omega$ and $T_2$.

\begin{figure}
\centering
  
  \includegraphics[width=.32\columnwidth]{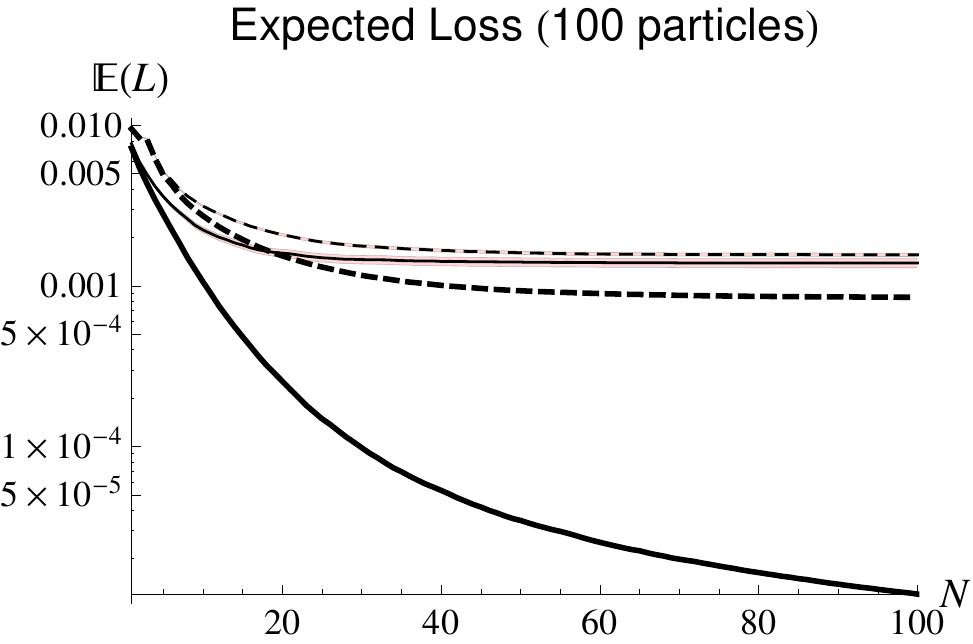}
  \includegraphics[width=.32\columnwidth]{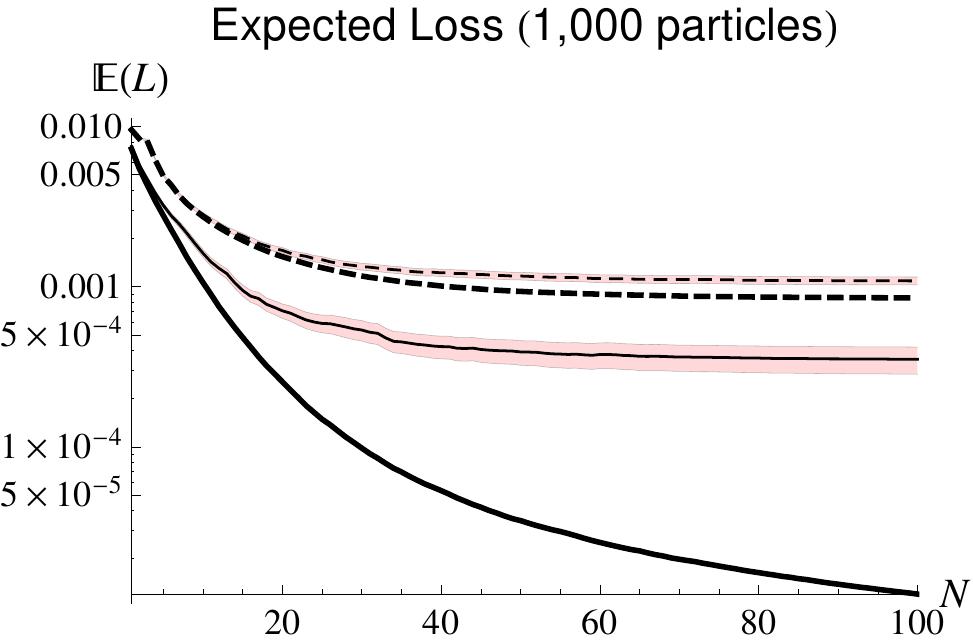}
  \includegraphics[width=.32\columnwidth]{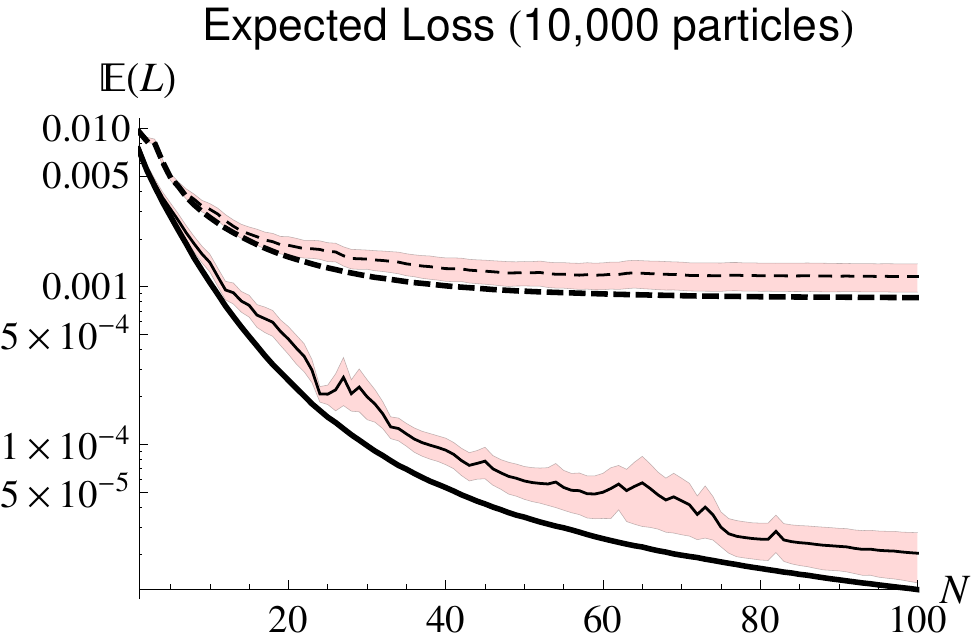}
  \vskip1em
  \includegraphics[width=0.8\columnwidth]{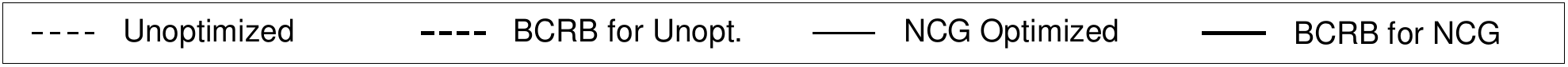}

  \caption{\label{fig:1Dscale} Left to right: the performance, as a function of the number of measurement $N$, of the sequential Monte Carlo algorithm for $n = 100, 1\; 000,$ and $10\; 000$ particles. The model is that of equation \eqref{SMC1paramModel} with $T_2=100\pi$ (see also figure \ref{fig:SMC1Dlike}).  The dashed lines indicate data taken without local optimization, while the solid lines indicate trials in which initial guesses were optimized using the NCG method. For each data set, the corresponding thick line indicates the Bayesian Cramer-Rao bound. Errors in estimating the performance are indicated by red shaded regions around each curve.}
\end{figure}

\subsection{Results for Known $T_2$ Model}
\label{sec:benchmarking-consistency}
Our first set of numerical experiments examines the performance of our algorithm for the case where 
the decoherence time is known with perfect precision.  Specifically, we take $T_2=100\pi$, $\omega\sim \mathcal{N}(0.5,0.01)$ and choose the experimental times to
be spaced uniformly such that the
$k^{\rm th}$ experiment occurs at time $t_k=2k\pi/3$ (chosen arbitrarily) and examine the mean--square error in $\omega$ averaged over a minimum of $1625$ trials with randomly chosen $\omega$.  
  This 
data is presented in Fig.~\ref{fig:1Dscale}.  The figure shows that the mean--square error for $\omega$
decreases as we vary the number of particles from $100$ to $10\ 000$, and in particular gives a relative MSE that is less than $1\%$ for $N\ge 100$.  We also see that the data
remain close to the BCRB (which is a lower bound on the MSE) for either $n=1\ 000$ or $10\ 000$ particles if no optimization is used, whereas 
$n=10\ 000$ is needed to approach the BCRB in the case where NCG is used.  This not only shows that several thousand particles should be sufficient for our purposes, but also that the MSE yielded by our algorithm scales near--optimally if a sufficiently large number of particles are used in the SMC approximation.

\begin{figure}[t]
  \centering
  \subfigure{\includegraphics[width=0.45\columnwidth]{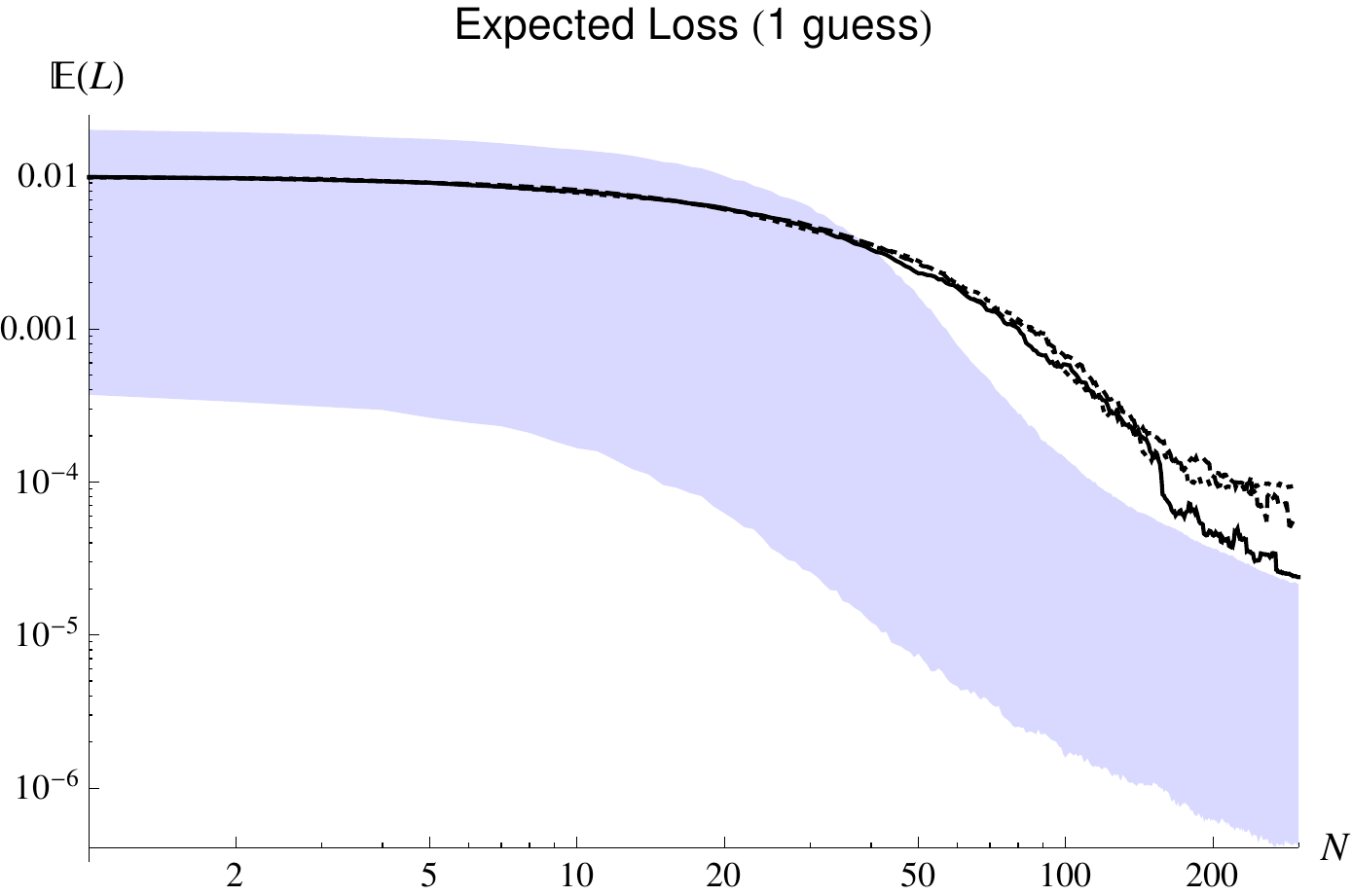}}
  \hskip3em
  \subfigure{\includegraphics[width=0.45\columnwidth]{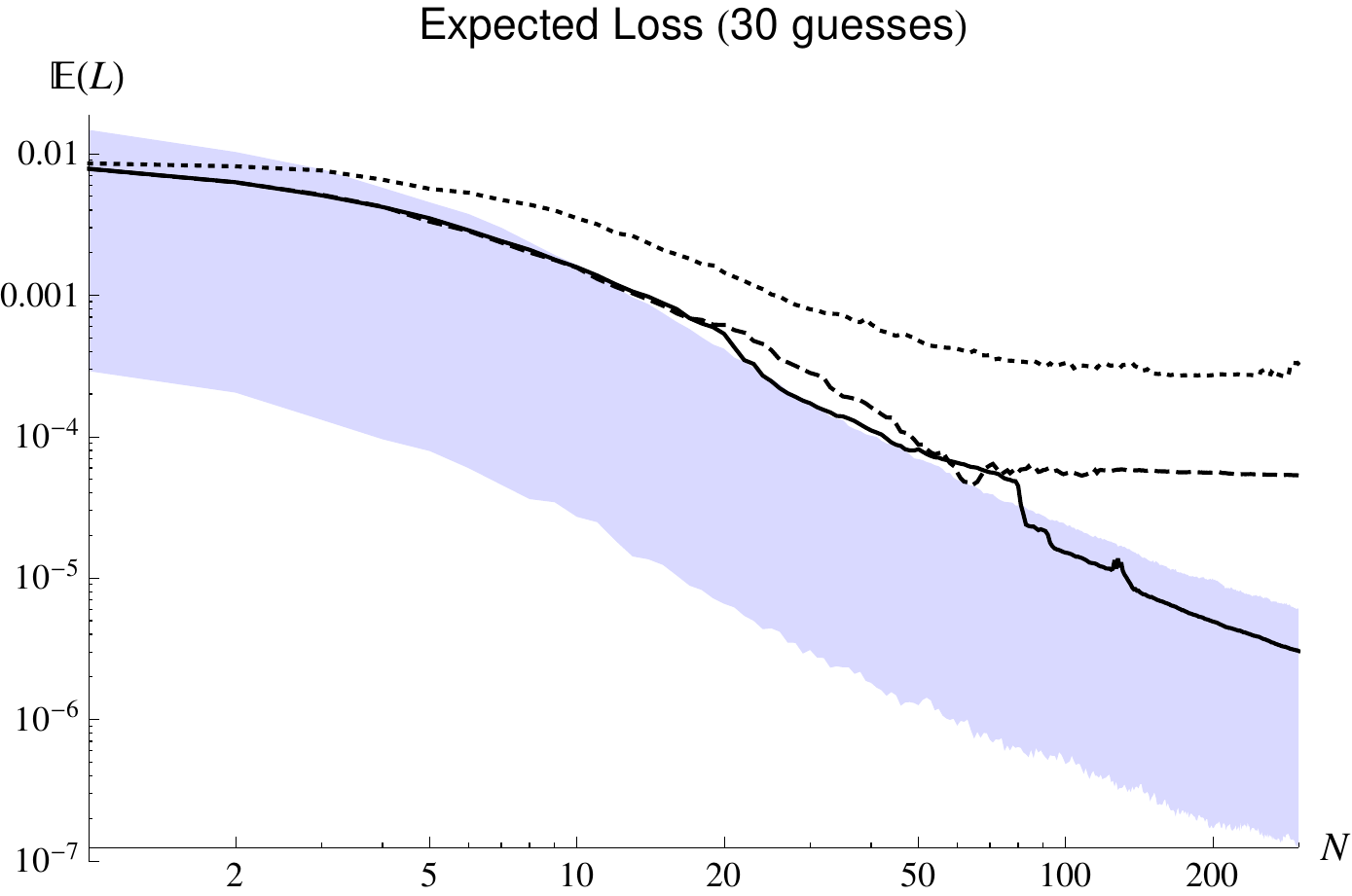}}
  \includegraphics[width=0.8\columnwidth]{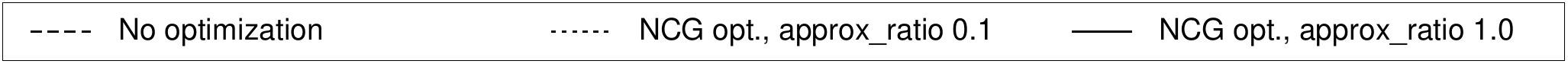}
\caption{This figure compares the mean--square error as a function of the number of experiments used for the known $T_2$ model with $T_2=100$, $5000$ particles and an approximation ratio of $1$ with guessed experimental times chosen randomly from an exponential distribution with mean $T_2$. On the left, data is shown for 1 guess, while on the right, we show data for 30 guesses. The shaded region in each plot indicates a $68\%$ confidence interval for data collected using NCG optimization with $\texttt{approx\_ratio} = 1.0$.
\label{fig:classicalcost}}
\end{figure}

Figure~\ref{fig:classicalcost} provides a more in depth analysis of the error scaling for the known $T_2$ model.  The previous
data set contained only one experimental guess per experiment, whereas in general we could consider making many guesses for the
optimal experiment and choosing the one with the highest utility $U(t) = -\expect_D[L]$.  We assess the performance of our algorithm as a function of the number of guesses by introducing a new guess heuristic.  Instead of choosing uniformly spaced guesses, we choose each guessed time randomly from an exponential distribution with mean $T_2$.  This guess heuristic also has the advantage of using very little intuition about the structure of the Hamiltonian that we are attempting to parameterize, which means that we expect the performance of this heuristic to be a better estimate of the worst--case performance of our algorithm.  

Unsurprisingly, we find that the MSE is reduced if we choose the best of $30$ possible experiments rather than a single randomly chosen experiment.
We also see that local optimization tends to improve the quality of the approximation if we pick the approximation ratio to be $1$.  Figure~\ref{fig:classicalcost} also shows that smaller values of the approximation ratio can cause the mean--square error to saturate at relatively large values.  In fact, taking $\texttt{approx\_ratio}=0.1$ was sufficient to cause the data taken for $30$ guesses and NCG optimization to have a \emph{larger} mean--square error in $\omega$ than the case with no optimization and $1$ random guess.  For this reason, we take the $\texttt{approx\_ratio}=1$ for most of the examples in this section. In Section \ref{sec:benchmarking-hyperregion-est}, we shall see an example where $\verb+approx_ratio+ < 1$ provides more benefit.

An individual trial  may have a mean--square error that  differs significantly from the expected loss.  The shaded regions in Fig.~\ref{fig:classicalcost} give $68\%$ confidence
intervals for the actual loss for the case with NCG and $\texttt{approx\_ratio}=1$ (the confidence intervals for the other data sets were similar), and find the surprising result that the mean--square error is frequently
outside the confidence interval in the cases where NCG optimization is not used.  This suggests that the distribution of the utility of experiments can wildly vary and that the distribution is skewed because a significant fraction of the guesses provide virtually no information.  NCG minimizes the chances of choosing an uninformative experiment and hence it forces the guesses towards the more informative experiments.  We should also note that there is room for improvement here, since the MSE is closer to the upper limit of the confidence interval. Sophisticated optimization procedures may therefore be of use when searching for informative experiments given an uninformed guess heuristic (such as our random guess heuristic).

These results show that poor guess heuristics can be mitigated using our algorithm by optimizing over more guesses and using local optimization of the experiments.  We also note that  the median square--error performance of our algorithm tends to be much better than the mean square--error because a small fraction of the trials randomly choose very bad guesses.  We see that NCG optimization can be used to cause the mean--square error to approach the median--square error.  We therefore find that estimates of the error (such as the posterior variance or region estimates of $\omega$) are needed in order to guarantee that  a particular trial is not pathological.

\subsection{Region Estimation}
\label{sec:benchmarking-region-est}
One of the most substantial contributions of our algorithm is its ability to provide region estimates for 
the location of the true Hamiltonian, which allow us to quantify our uncertainty in the true model parameters.
We compare the probability mass enclosed by the covariance region estimator described in Section \ref{sec:region-est}. 
  A simplifying assumption is made in our analysis: we assume that the posterior distribution
is approximately Gaussian.  Although difficult to justify theoretically, we have yet to find an example for the models considered
here where the posterior does not appear Gaussian after a sufficiently large number of experiments.  Under the Gaussian model
of the posterior distribution, we expect the true model parameters to be within an ellipse described by the covariance matrix whose
volume is then described by the $Z$--score used.  For example, in the one--dimensional case approximately $95\%$ of the probability
mass is located within $2$--standard deviations, which corresponds to $Z=2$.  We choose $Z=3$ standard deviations
from the mean for these examples which correspond to probability masses of 
$\tilde{m}(\Cov(\hat{\vec{x}})^{-1}/Z^2) \approx 0.9973$ and $\tilde{m}(\Cov(\hat{\vec{x}})^{-1}/Z^2) \approx 0.9946$ for the 
one-- and two--parameter cases respectively.

\begin{figure}
  \begin{center}

\includegraphics[width=0.6\columnwidth]{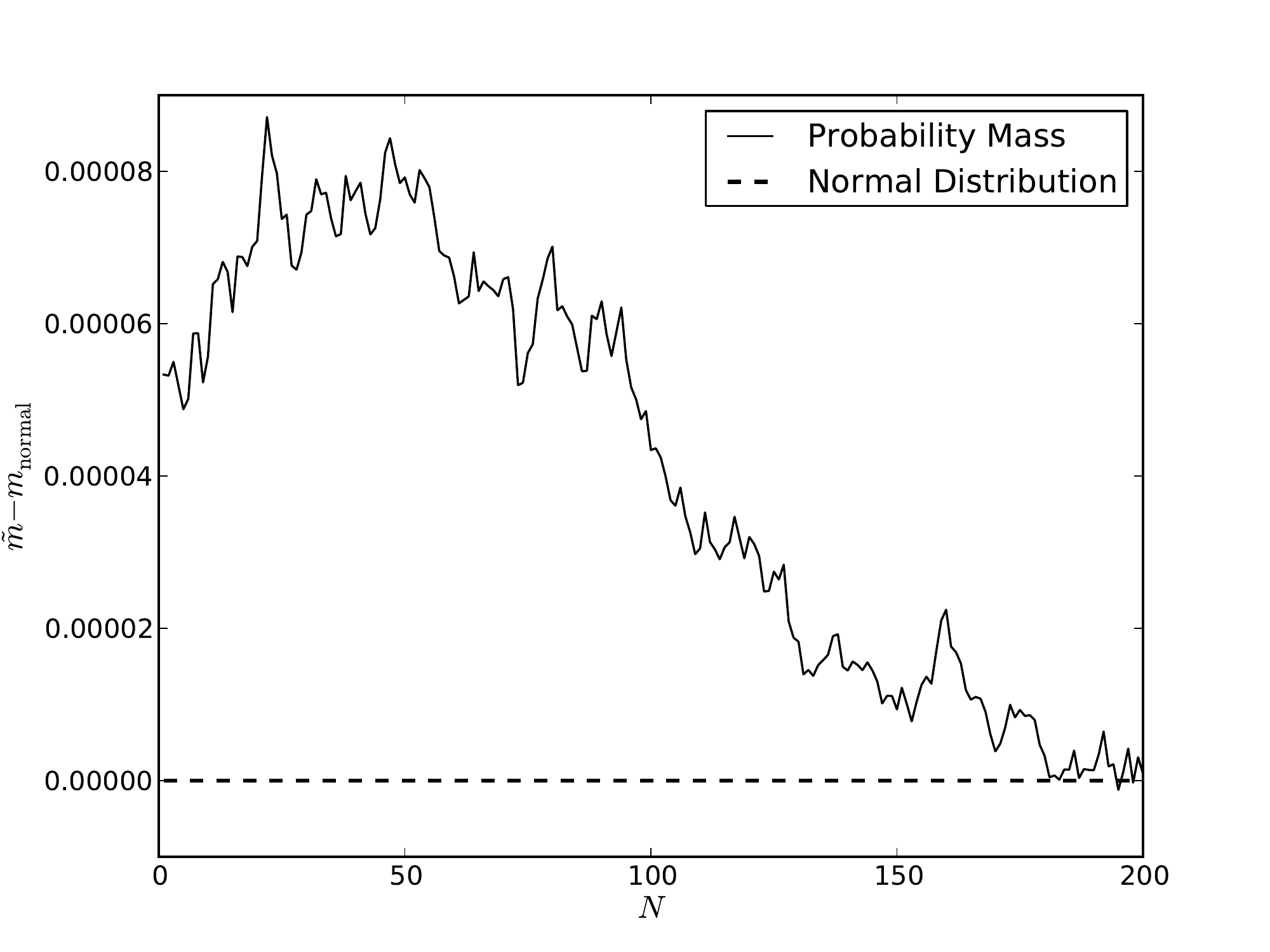}

    \caption{Sequential Monte Carlo approximated covariance probability mass $\tilde{m}(\Cov(\hat{\vec{x}})^{-1}/Z^2)$ for the known-$T_2$ single-parameter model as compared to the probability mass $m_{\text{normal}} \approx 0.9973$ expected for the normal distribution and as function of the number $N$ of experiments
      performed, averaged over 20\ 119 trials using a guess heuristic that chooses the $k^{\rm th}$ guess to occur at time $(9/8)^k$, using $30$ guesses, $1\ 000$ particles and NCG optimization. The dashed line shows the probability mass for the corresponding normal distribution.
  \label{fig:cov-prmass-sq_T2}}

  
    %

  \end{center}
\end{figure}

Figure \ref{fig:cov-prmass-sq_T2} illustrates that the approximate probability mass $\tilde{m}$ approaches
the probability mass we would expect for a normal distribution for the known-$T_2$ model (introduced in Section \ref{sec:single param}) in the limit of large $N$, providing evidence in favor of our use of the covariance ellipse as a region estimator on the posterior. In particular, we note that the value of $\tilde{m}$ approaches $0.9973$, such that the quality of the Gaussian approximation improves as we collect data.
The transient behavior for small experiment numbers occurs because insufficient experiments have been considered for the posterior to approach a Gaussian. In this specific example, the average differences in enclosed probability mass after each experiment are on the order of $0.01\%$, and thus may not be of practical signifigance.

\subsection{Results for Unknown $T_2$ Model}
\label{sec:benchmarking-T2}
We now turn our attention to the  comparably challenging task of learning Hamiltonian parameters
without a precise estimate of $T_2$.  These calculations were performed using
the true distributions $\omega \sim \mathcal{N}(0.5, 0.0025)$ and $1/T_2 \sim \mathcal{N}(0.001, 0.00025^2)$, and with
the scale matrix $\matr{Q} = \diag(1, 0.0025/0.00025^2)=\diag(1,100)$.  The guess heuristic that we focus on chooses
times randomly from an exponential distribution with mean $1\ 000$, corresponding to the mean value of $T_2$ according to the initial prior.

\begin{figure}[t!]
  \begin{center}

  \subfigure[\label{fig:unkT2-mse-unopt-omega}] {\includegraphics[width=0.45\columnwidth]{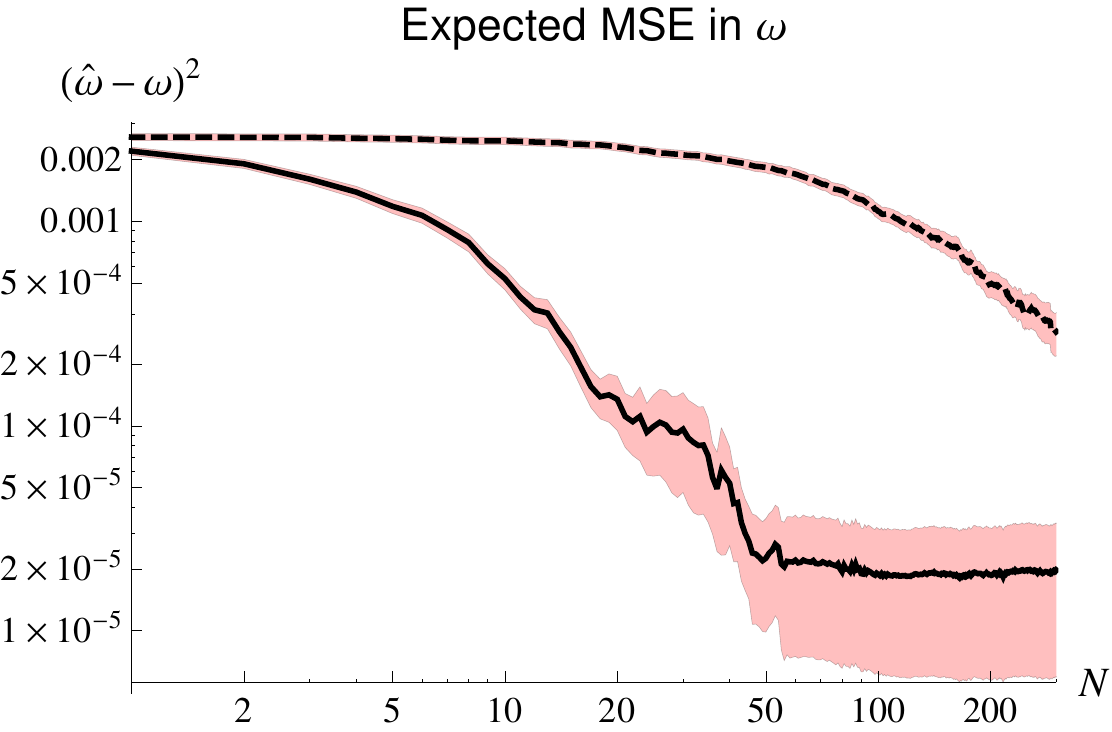}}
  \subfigure[\label{fig:unkT2-mse-unopt-gamma2}]{\includegraphics[width=0.45\columnwidth]{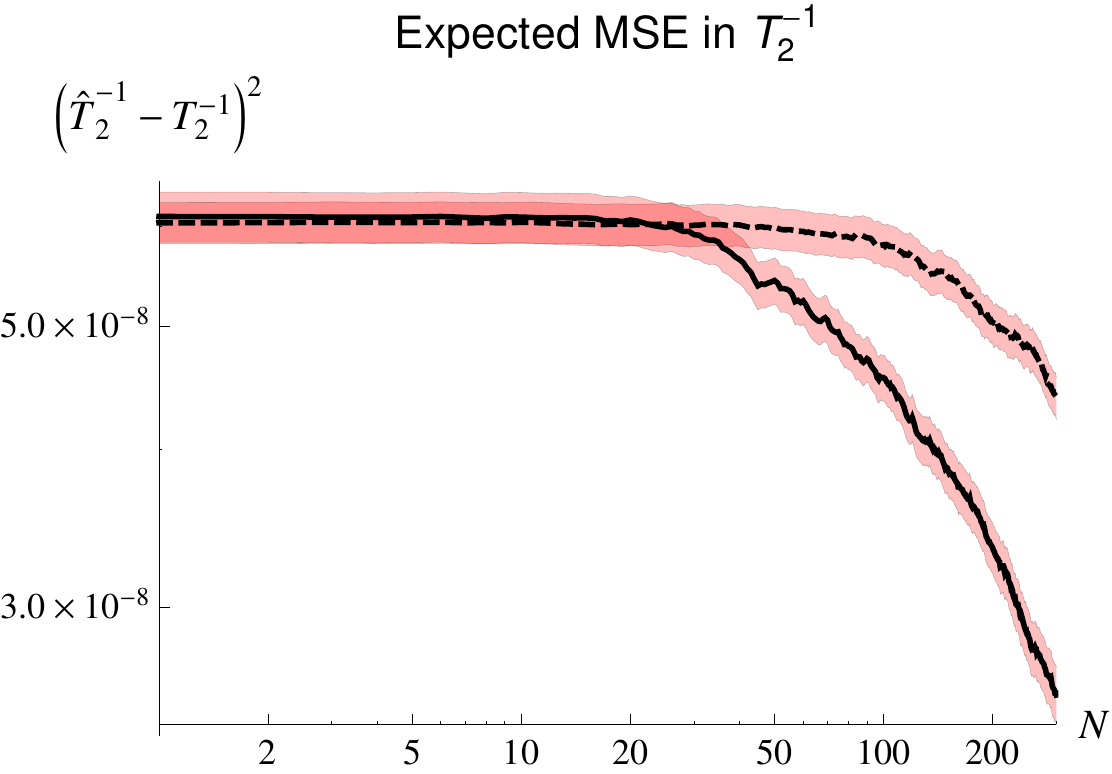}}
  \includegraphics[width=0.6\columnwidth]{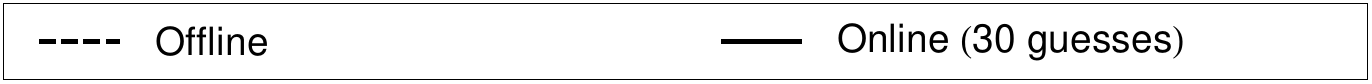}

  \caption{\label{fig:unkT2-performance-unopt} Benchmarking of the ``unknown-$T_2$''model (Section \ref{sec:two param}) using $n = 5\; 000$ particles and random initial guesses without local optimization. Data indicated dashed lines correspond to trials where a single initial guess was used for each experiment, while data indicated by solid lines were collected using 30 guesses per experiment. The single-guess data is averaged over 1,380 trials while the 30-guess data is averaged over 1,109 trials. Errors in estimating performance are indicated by red shaded regions about each curve.}
  
  \end{center}
\end{figure}

We examine the variation of the MSE with the number of guesses used in Fig.~\ref{fig:unkT2-performance-unopt}. The figure shows that, in the absence of local optimization of experiment times, the MSE for both $\omega$ and $1/T_2$ is significantly improved by using an increased number of guesses.  In particular, we find that if $30$ guesses are used, then only $50$ experiments are required on average to learn $\omega$ within a $0.9\%$ error, even without a well characterized $T_2$.  The improvement is much more substantial for $\omega$ than it is for $1/T_2$ because the contrast on $T_2$ is much less significant.

\begin{figure}[h!t]
  \begin{center}

  \subfigure[\label{fig:unkT2-mse-opt-omega}] {\includegraphics[width=0.45\columnwidth]{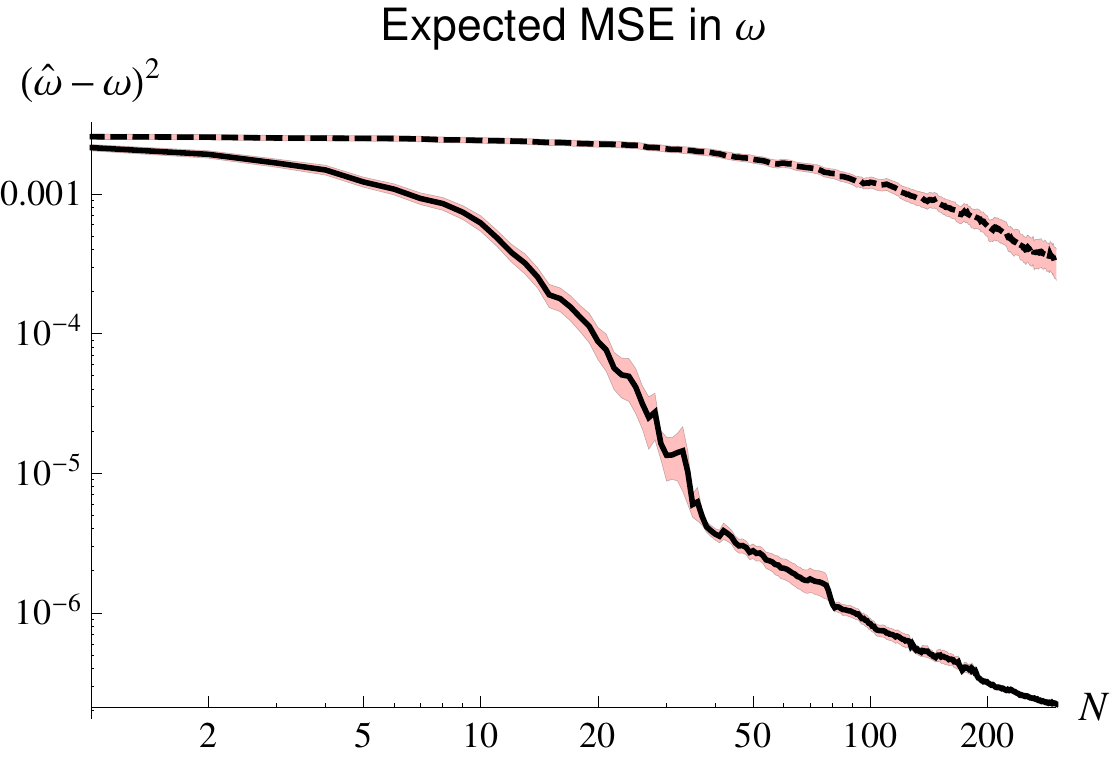}}
  \subfigure[\label{fig:unkT2-mse-opt-gamma2}]{\includegraphics[width=0.45\columnwidth]{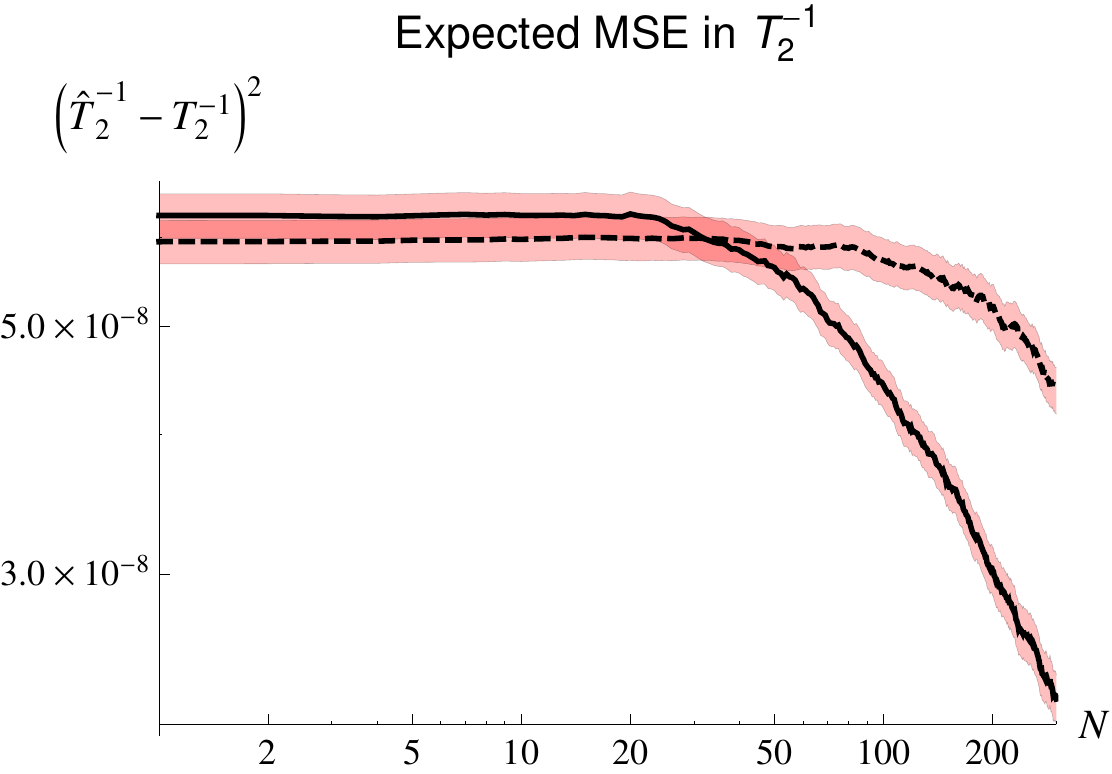}}
  \includegraphics[width=0.6\columnwidth]{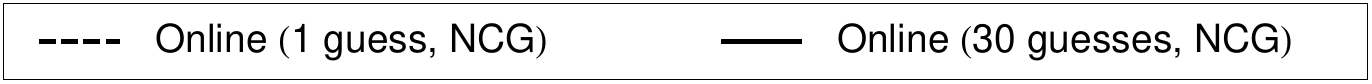}

  \caption{\label{fig:unkT2-performance-opt} Benchmarking of the ``unknown-$T_2$''model (Section \ref{sec:two param}) using $n = 5\; 000$ particles and random initial guesses with local optimization by the NCG method. Data indicated dashed lines correspond to trials where a single initial guess was used for each experiment, while data indicated by solid lines were collected using 30 guesses per experiment. The single-guess data is averaged over 1,023 trials while the 30-guess data is averaged over 930 trials. Errors in estimating performance are indicated by red shaded regions about each curve.}
  
  \end{center}
\end{figure}

\begin{figure}[t]
  \begin{center}

  \subfigure[\label{fig:unkT2-mse-opt_v_unopt-omega}] {\includegraphics[width=0.45\columnwidth]{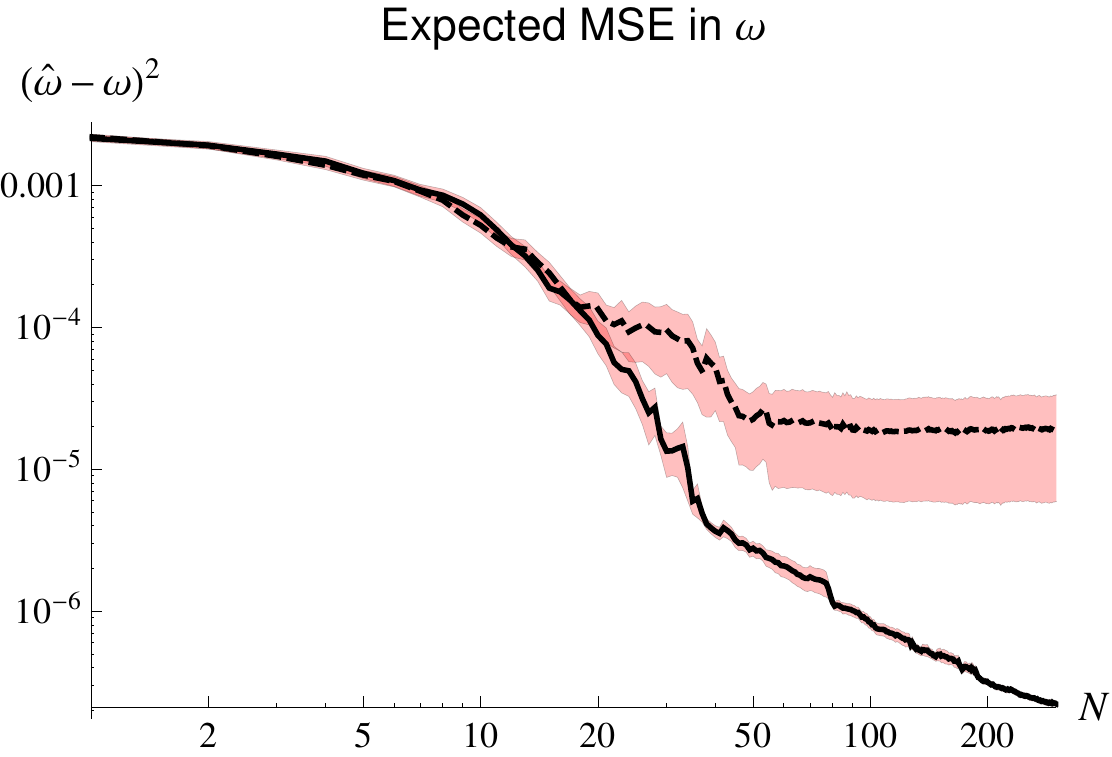}}
  \subfigure[\label{fig:unkT2-mse-opt_v_unopt-gamma2}]{\includegraphics[width=0.45\columnwidth]{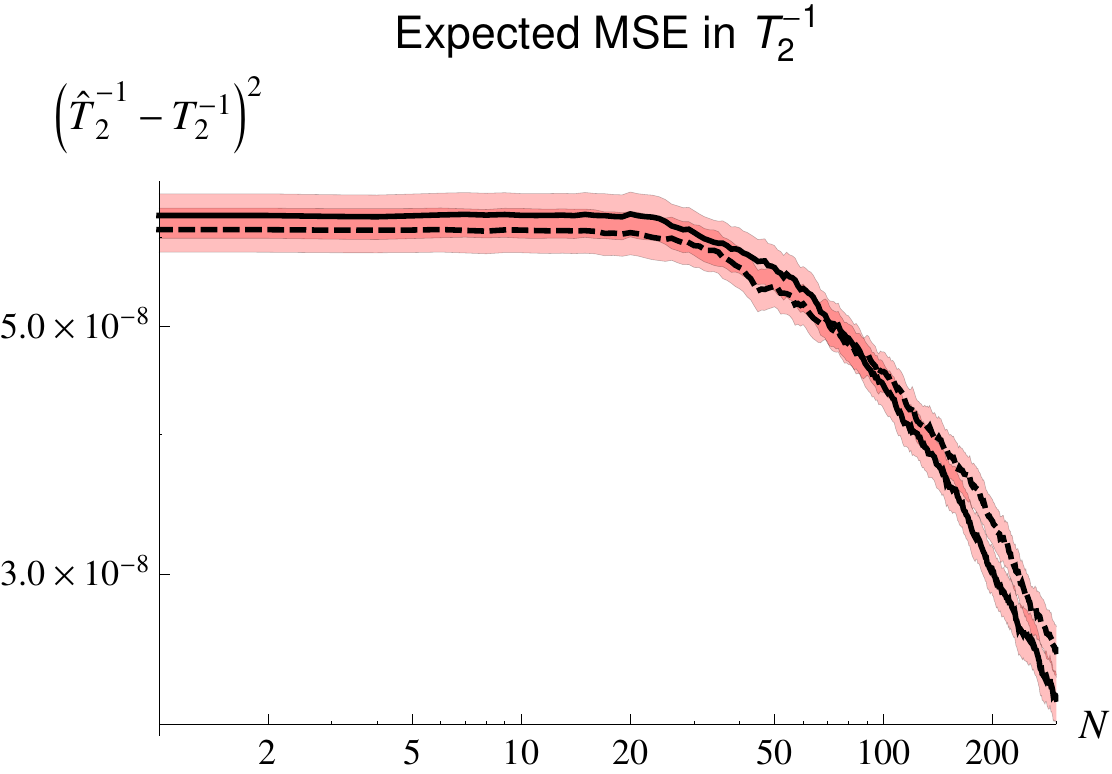}}
  \includegraphics[width=0.6\columnwidth]{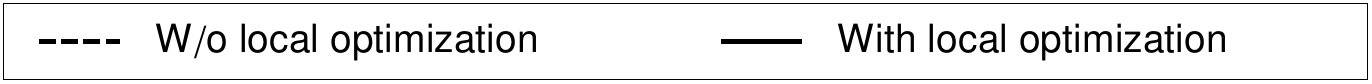}

  \caption{\label{fig:unkT2-performance-opt_v_unopt} Benchmarking of the ``unknown-$T_2$''model (Section \ref{sec:two param}) using $n = 5\; 000$ particles and 30 random initial guesses. Data indicated dashed lines correspond to trials where a each initial guess was used without local optimization, while data indicated by solid lines were collected using NCG optimization for each guess. The unoptimized data is averaged over 1,109 trials while the optimized data is averaged over 930 trials. Errors in estimating performance are indicated by red shaded regions about each curve.}
  
  \end{center}
\end{figure}

Fig.~\ref{fig:unkT2-performance-opt} examines the effect of increasing the number of guesses for strategies that use NCG.
The most significant qualitative difference  between the data collected using NCG and that of Fig.~\ref{fig:unkT2-performance-unopt}
is that the MSE for $\omega$ shows no evidence of saturating and instead continues to shrink as the number of experiments are increased (as seen most clearly in Fig.~\ref{fig:unkT2-performance-opt_v_unopt}).  This implies that our randomized guess heuristic is unlikely to randomly guess very informative experiments after a fixed number of experiments, but the landscape is sufficiently devoid of local optima that NCG optimization finds informative experiments in the vicinity of our uninformed guesses.  We also observe that NCG does not substantially improve the MSE if $1$ guess is used.  This suggests that the landscape is not sufficiently convex that local optimization about an individual guess is likely to find experiments that are substantially more informative.
We therefore conclude, again, that increasing number of guesses used and using NCG substantially improves the MSE for $\omega$ and has a much more subtle effect on the
knowledge of $T_2$ if  local optimization is used.  

\begin{figure}
	\centering
	\subfigure[\label{subfig:unkT2-estquality-omega}]{\includegraphics[width=.45\columnwidth]{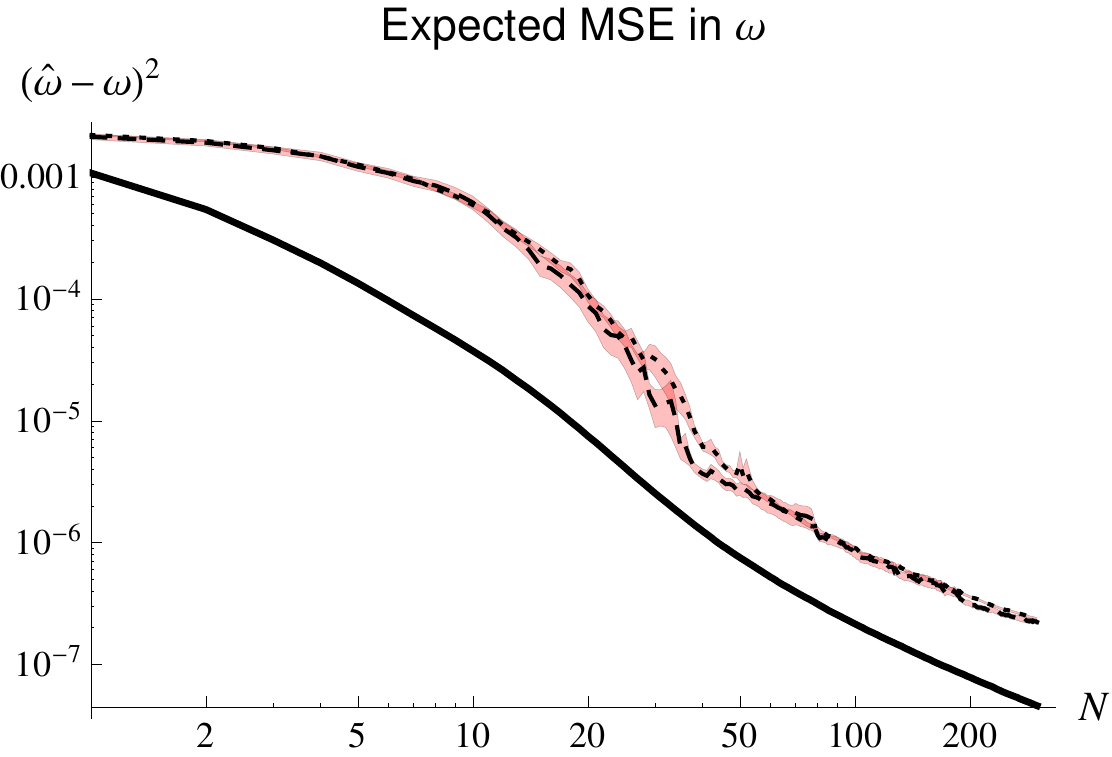}}
	\subfigure[\label{subfig:unkT2-estquality-T2}]{\includegraphics[width=.45\columnwidth]{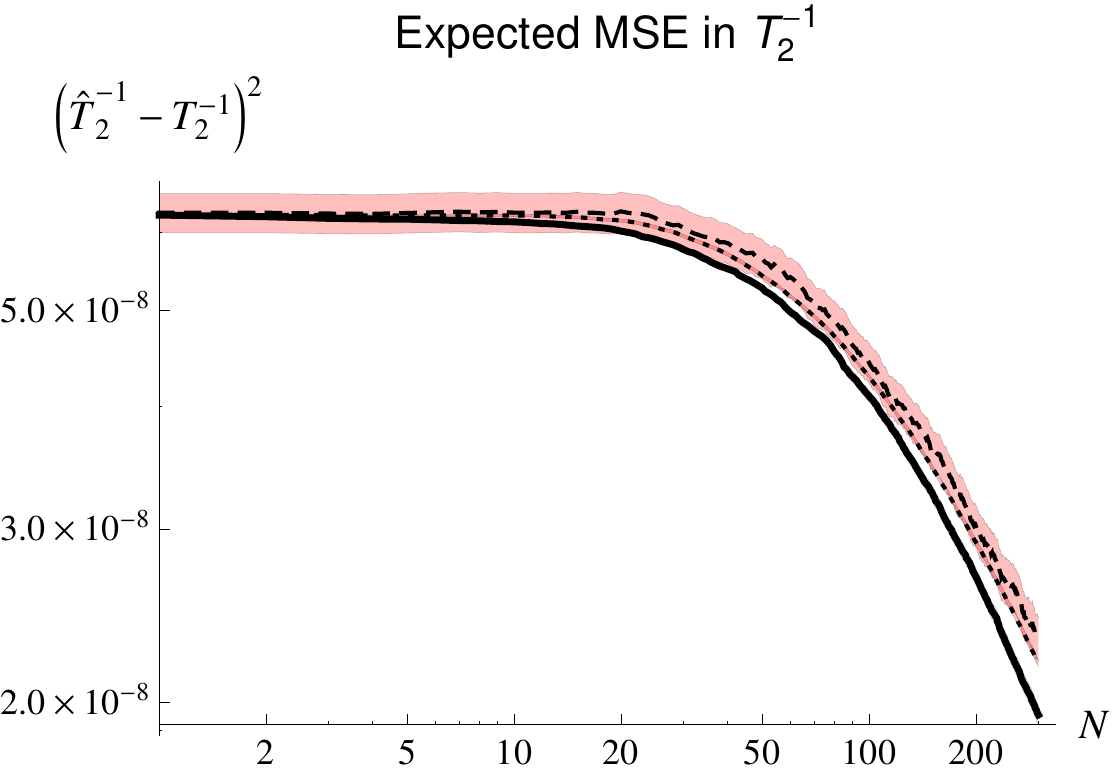}}
	
	\includegraphics[width=.55\columnwidth]{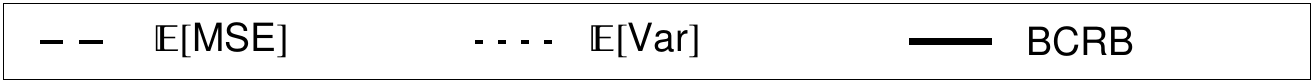}

	\subfigure[\label{subfig:unkT2-estquality-Q}]{\includegraphics[width=.45\columnwidth]{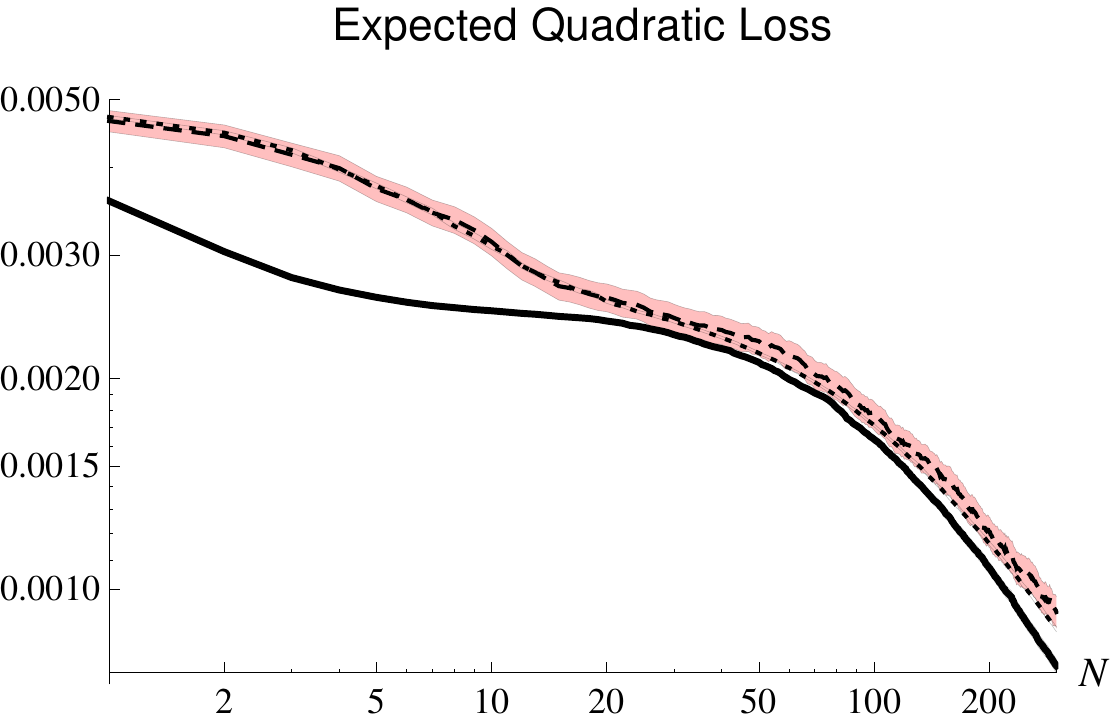}}
	
	\includegraphics[width=.55\columnwidth]{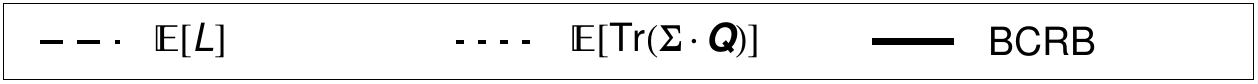}
	\caption{\label{fig:2Dscale} The actual and estimated performance, as a function of the number of measurements $N$, of the sequential Monte Carlo algorithm for $n = 5\; 000$ particles. The model is that of equation \eqref{SMC2paramModel} with unknown $T_2$ (which is estimated as $\Gamma = 1/T_2$ for numerical precision considerations).  The dotted curve is the posterior variance of the particles; dashed is the actual mean squared error and solid is numerically calculated Bayesian Cramer-Rao lower bound. In subfigures (a) and (b), the MSE and variances are those of the individual parameters $\omega$ and $T_2^{-1}$, respectively, while subfigure (c) shows the actual and estimated quadratic losses scaled using $\matr{Q}=\diag(1, \sigma^2_{\omega} / \sigma^2_{T_2^{-1}})$, where $\sigma^2_{\omega}$ and $\sigma^2_{T_2^{-1}}$ are the variances in $\omega$ and $T_2^{-1}$ according to the initial prior $\pi$.}
\end{figure}

Similarly to the case of known $T_2$, it is useful to benchmark the performance of our algorithm against the BCRB, which gives a lower bound on the  MSE.  Figure~\ref{fig:2Dscale} provides a comparison of the MSE, the estimate of the MSE given by the variance of the posterior and the BCRB for $\omega$, $T_2^{-1}$ and ${\rm Tr}(\Sigma\cdot\matr{Q})$.  We see that the expected posterior variance is typically within statistical error of the MSE for all three of these quantities, suggesting that the posterior variance can be used as a very good estimate of the MSE for this model.  We also note that the MSE is very close to the MSE for the $T_2^{-1}$ data and ${\rm Tr}(\Sigma\cdot\matr{Q})$.  The MSE for $\omega$ is within a constant multiple of the BCRB.  We do not, in fact, expect that the MSE in $\omega$ should approach the BCRB because the algorithm chooses experiments to optimize ${\rm Tr}(\Sigma\cdot\matr{Q})$ rather than the error for either $\omega$ or $T_2^{-1}$ individually.

\subsection{Hyperparameter Region Estimation Performance}
\label{sec:benchmarking-hyperregion-est}

Having demonstrated the effectiveness of our region estimation algorithm, it remains to show that the generalization to hyperparameter regions works as described in Section \ref{sec:hyperparam-est}. The objective here is to analyze the robustness of our algorithm in the presence of fluctuating ``true'' parameters of the Hamiltonian.
We do so by using the Gaussian hyperparameter model of Equation \eqref{eq:hyperparameter-model-gaussian} as discussed in Section \ref{sec:hyperparameter-model}, then comparing the model parameter region volume and probability mass for the region estimated from Equation \eqref{eq:hyperregion-cov} to the volume and probability mass of the corresponding ``true'' model parameter region. We benchmark this model by choosing ``true'' hyperparameters $\mu$ and $\sigma^2$ for $\omega$ according to the normal distribution
\begin{align}
  \mu, \sigma^2 \sim \mathcal{N}\left[(\mu_{\mu}, \mu_{\sigma^2}), \diag(\sigma^2_{\mu}, \sigma^2_{\sigma^2}) \right].
\end{align}
Recall that the unknown frequency is distributed as $\omega\sim \mathcal{N}(\mu,\sigma^2)$.
In particular, this true distribution does not admit any correlation between the mean and variance hyperparameters. We then use the true distribution as our prior distribution.

\begin{figure}
  \begin{center}

    \subfigure[\ Probability mass contained by region estimators. \label{fig:hyperregion-est-prmass}]{
      \includegraphics[width=0.45\columnwidth]{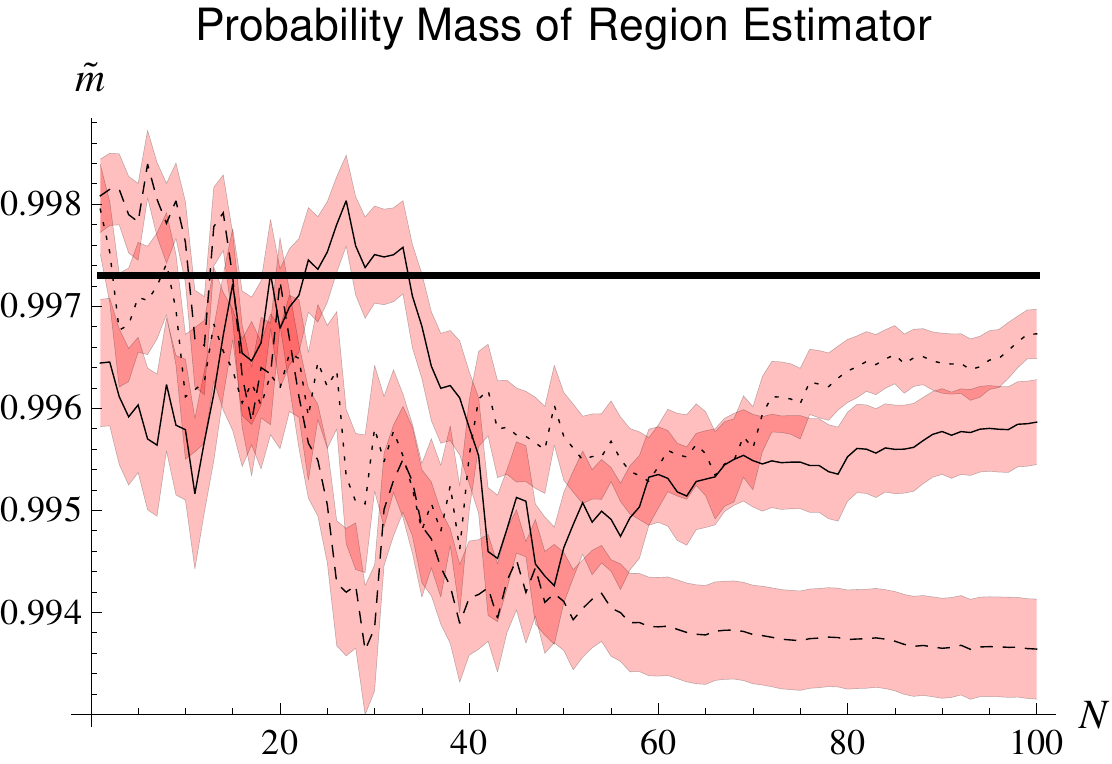}
    }    
    \subfigure[\ Comparison of estimated and true model variances. \label{fig:hyperregion-est-cov}]{
      \includegraphics[width=0.45\columnwidth]{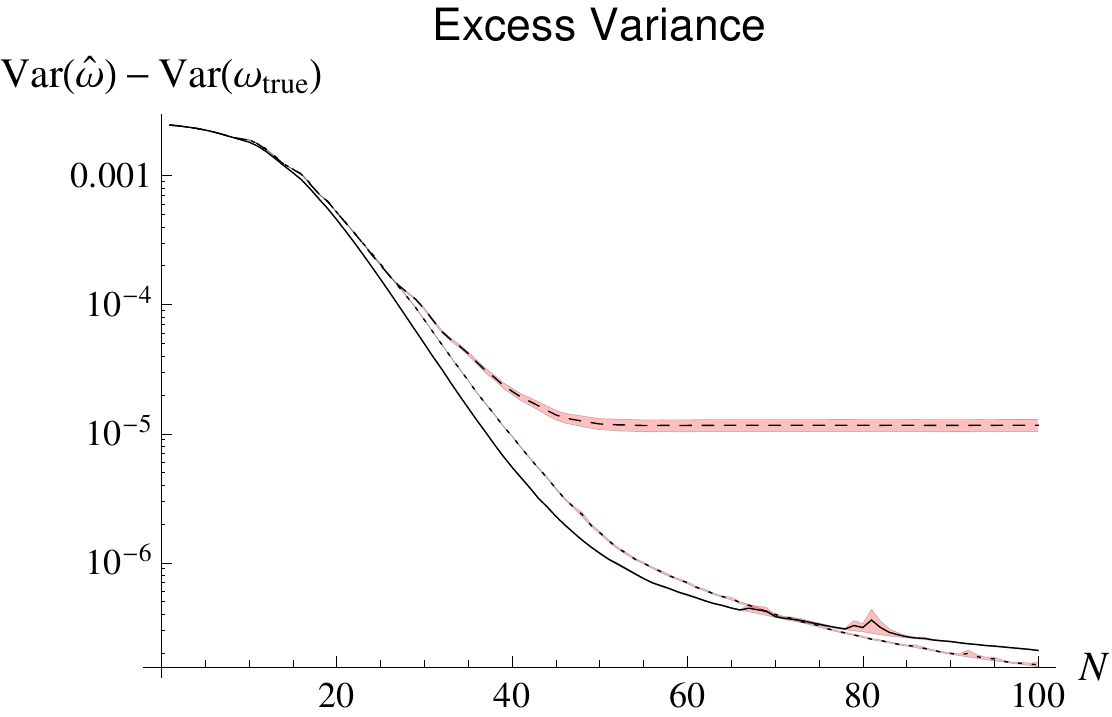}
    }
    
    \includegraphics[width=0.55\columnwidth]{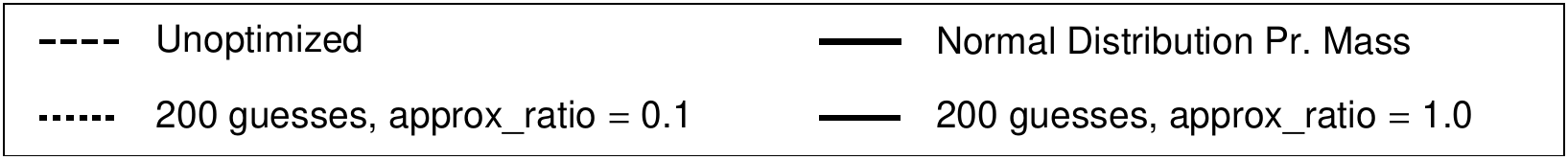}

  \caption{Benchmarking region estimators for Gaussian hyperparameter model (Section \ref{sec:hyperparameter-model}) using $n = 2\; 000$ particles, $\omega\sim\mathcal{N}(\mu,\sigma^2)$ where $\mu\sim \mathcal{N}(0.5,0.001^2)$ and $\sigma^2\sim \mathcal{N}(0.0025,0.0025^2)$.}
  
  \end{center}
\end{figure}

Figure~\ref{fig:hyperregion-est-prmass} provides estimates of the probability mass contained within our estimated region for the Hamiltonian, which uses a $Z$-score of $3$.  Since we assume a Gaussian posterior, we anticipate that $99.7\%$ of the probability mass should lie within the region estimation of $\expect[\hat{\omega}] \pm 3 \sqrt{\Var(\hat{\omega})}$.  We find very good agreement with this assumption, and find that at worst $99.4\%$ of the probability mass for the hyperparameters lies within the estimated region.  The data also suggests that these small differences vanish for the optimized data sets, which appear to approach the ideal enclosed probability mass of $99.7\%$ in the limit of large $N$.  It is also interesting that the data with $\texttt{approx\_ratio}=0.1$ yielded the best region estimation for the probability mass (unlike the previous examples).  This is likely because the low approximation ratio de--emphasizes the tails of the posterior and non--Gaussian behavior typically is manifested in the tails.  This shows that there are, perhaps surprisingly, examples where taking $\texttt{approx\_ratio}<1$ is useful.

Hyperparameters are not typically a quantity of interest by themselves.  They usually are of relevance because they parameterize a distribution of the unknown parameter.  Following Equation \eqref{eq:hyperregion-var}, we calculate $\Var(\hat{\omega})$ as
\[
    \Var(\hat{\omega}) = \Var(\hat{\mu}) + \expect[\hat{\sigma}^2].
\]
Figure \ref{fig:hyperregion-est-cov} compares $\Var(\omega)$ to the variance parameter $\sigma^2$.  As the number of experiments grows, our region estimator for $\omega$ slightly overestimates the ``true'' variance of $\omega$ (on average).  We see from Fig.~\ref{fig:hyperregion-est-cov} that the estimatate of the variance of the unknown frequency that is inferred from our region estimate of the hyperparameters systematically over--estimates the variance of $\omega_{\rm true}$ on average.  This bias vanishes as the number of experiments increases.  We can therefore conclude that we can use the method of hyperparameters to robustly estimate the distribution of an unknown frequency, even in the presence of noise.

\subsection{Computational Cost}
\label{sec:benchmarking-classical-cost}
Another way that we can assess the cost of inferring the Hamiltonian of a system is in terms of the classical computing time needed to 
learn the Hamiltonian parameters to within a fixed error tolerance (as measured by the number of likelihood calls made).  Our previous discussion found that the experimental
time (measured by the number of experiments) can be minimized by choosing measurements that minimize the risk, and showed that increasingly sophisticated heuristics for generating these guesses tended to reduce the experimental time.  This suggests that a trade-off may be present between the experimental time and the classical processing time needed to learn the parameter.  This tradeoff will become increasingly relevant as the size of the quantum system grows, since existing quantum simulation techniques do not scale efficiently with the number of particles in the system and thus the cost of performing a likelihood call may asymptotically become much more expensive than performing an experiment.

\begin{figure}[t]
  \centering
  \subfigure{\includegraphics[width=0.45\columnwidth]{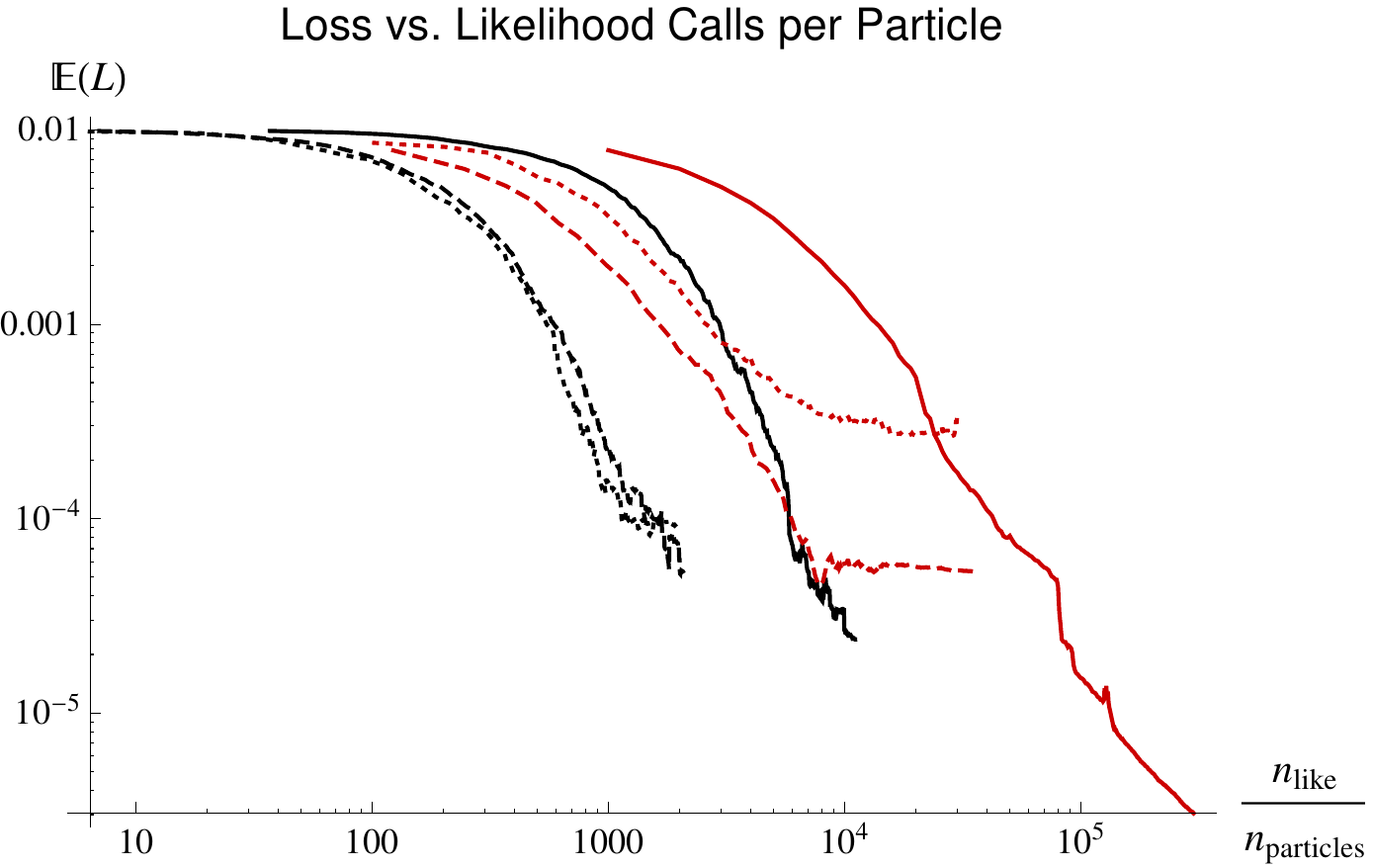}}
  \hskip3em
  \subfigure{\includegraphics[width=0.45\columnwidth]{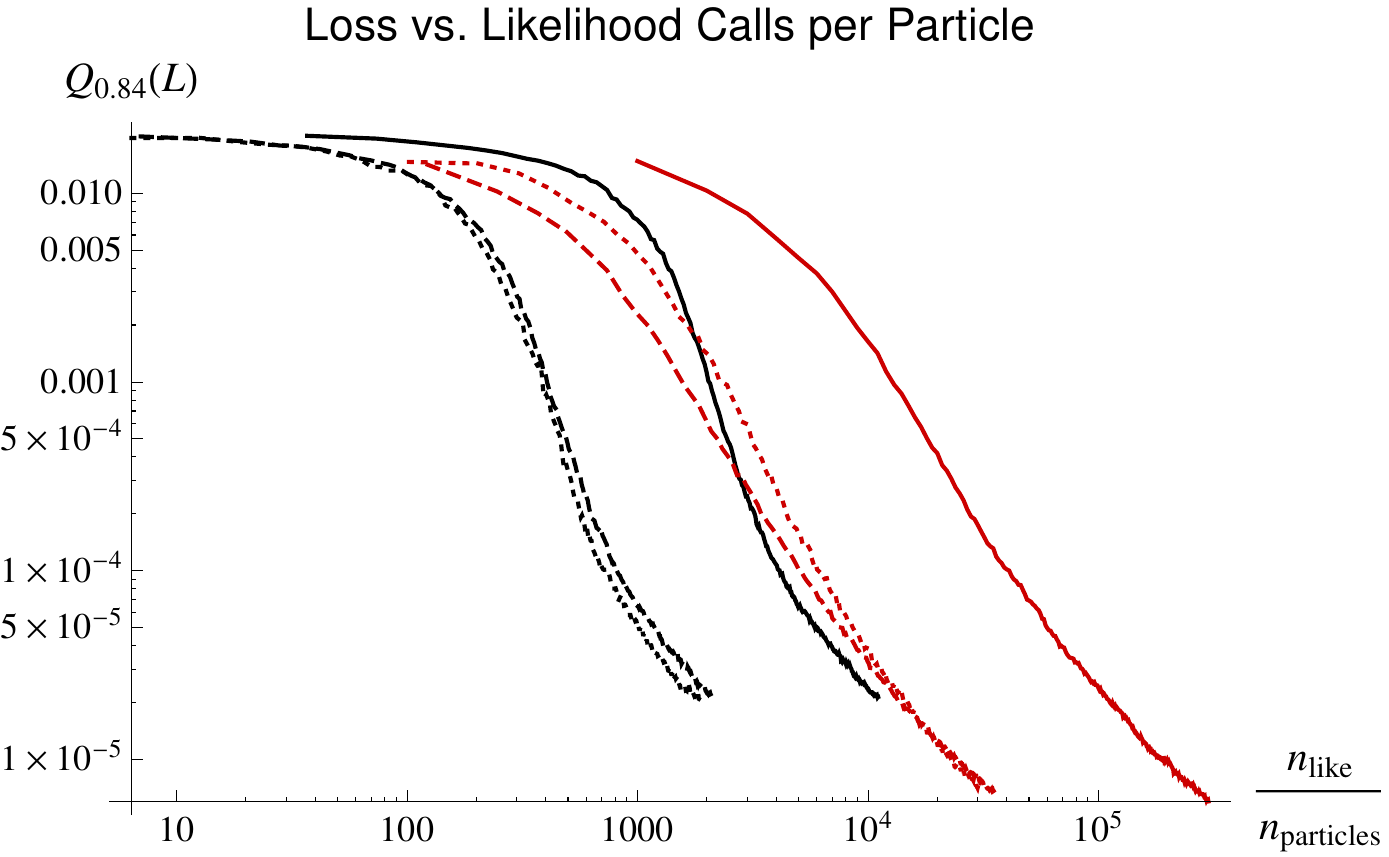}}
  \includegraphics[width=0.9\columnwidth]{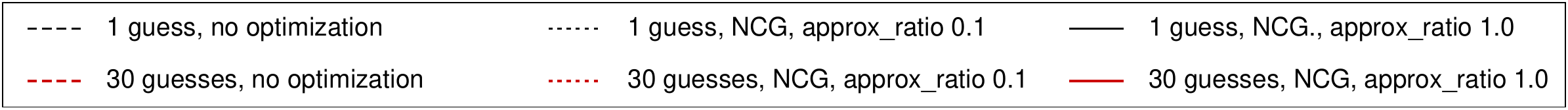}
  \caption{This figure compares the mean--square error as a function of the computational time for the known $T_2$  model with $T_2=100$, $5000$ particles,$\texttt{approx\_ratio}=1$ and guessed experimental times chosen randomly from an exponential distribution with mean $T_2$. The expected loss
  incurred by each optimization strategy is shown is shown in the left figure and the figure on the right shows the $84^{\rm th}$ percentile $Q_{0.84}$ of the loss, such that no more than $16\%$ of trials incur loss greater than the shown percentile. 
  \label{fig:classicalcost2}}
\end{figure}

If computational time is of primary importance (rather than experimental time), then the relative merits of the experimental design
heuristics changes.  In total, our data sets in figures~\ref{fig:classicalcost} and~\ref{fig:classicalcost2} required (on average) a number of likelihood calls that fell within the range $[1.05\times 10^{7},1.5\times 10^{9}]$.  A likelihood call required the evaluation of $\exp({-{t}/{T_2}})\cos^2\left(\frac\omega 2 t\right)+({1-\exp({-{t}/{T_2}})})/2$, which required time on the order of $10^{-7}$ seconds on our computers and lead to total computational times that were on the order of a second to a minute.  If the rate at which experiments can be performed were much faster than $200$ Hz then the utility of our algorithm as a means to speed up data collection may be lost.  If the two rates are approximately comparable, then interesting trade-offs appear between the computational time needed and the total experimental time.

These trade-offs become apparent by plotting the scaling of the MSE as a function of the computational time for the randomized guess heuristic in Fig.~\ref{fig:classicalcost2}.  The first feature that is obvious from the plot is that the strategies which yielded the lowest MSE per experiment tend to yield the highest MSE per likelihood call; although several of these strategies cause the expected loss (mean--square error) to saturate after a finite number of experiments.  In particular, this causes the strategy with $30$ guesses and no optimization as well as the strategy with $30$ guesses, NCG optimization and $\texttt{approx\_ratio}=0.1$ to intersect the curve for the cases with NCG optimization and $\texttt{approx\_ratio}=1$.  On the surface, this seems to indicate that the more expensive heuristics may have an advantage if small loss is desired; but this is misleading and to get a complete picture we need to look at more than just the expected performance of the strategies.

We can get a better understanding of this saturation by looking at the plot of the $84^{\rm th}$ percentile of the loss in Fig.~\ref{fig:classicalcost}, which shows that all of these strategies continue to provide improved estimates of $\omega$ even into this regime of saturation for at least $84\%$ of the trials considered.  This shows that there were a few trials where very poor guesses were chosen and the algorithm became stuck at a large MSE.  The data also suggests that the use of NCG and a large value of the approximation ratio can mitigate these problems, causing the learning algorithm to become more stable at the price of requiring more computational time.

There are many strategies that can be employed to reduce the computational time required by our algorithm.
Firstly, since the calculation of each guess is independent of the other guesses, this task can be trivially paralellized on a cluster that uses very little communication between the nodes.  Additionally, because
the simulations do not need to be high precision for the method to be successful, a single--precision implementation of the simulation step could also be used to improve the performance of the simulation in circumstances where the quantum simulation used to evaluate the likelihood function is computationally expensive. Such simulations have been demonstrated using graphical processing units (GPUs) \cite{gutierrez_parallel_2008} and field-programmable gate arrays (FPGAs) \cite{khalid_fpga_2004}; the latter is of particular interest due to the use of FPGA devices in the control of quantum information processing systems.

\section{Conclusions}
\label{sec:conclusion}

Our work provides a simple algorithm that applies Bayesian inference to
learn a Hamiltonian in an online fashion; that is to say, that our algorithm learns the Hamiltonian
parameters as the experiment proceeds rather than collecting data and inferring the Hamiltonian through post--processing.
This eliminates the need to store and process gigabytes of data that are recovered from even relatively short experiments.
Our work has several advantages over existing approaches to learning Hamiltonian parameters.
First, it can be used to estimate the optimal parameterization of the dynamics of an arbitrary 
quantum system within a space of model Hamiltonians.  Second, it can be used to provide
a region estimatate of the Hamiltonian parameters.  The importance of this is obvious: it allows us to
not only learn the unknown parameters but also quantify our uncertainty in them.
Third, our analysis of the algorithm shows a clear trade off between the experimental time and the computational
time needed to parameterize the Hamiltonian.  

We illustrated these advantages by benchmarking our algorithm's performance for a number of computationally tractable Hamiltonian models that involve a qubit precessing with an unknown frequency in the presence of decoherence (finite $T_2$ time).  Our results showed
that the scaling of the mean--square error of an unknown frequency with the number of experiments
irrespective of whether $T_2$ is known approaches the Bayesian Cramer-Rao bound, which is known to be optimal, given a set of experimental designs.  In contrast, other methods
for learning an unknown frequency require a well characterized $T_2$ time as a prerequisite.  Our work, on the other hand, shows that we can
learn the unknown frequency and the unknown $T_2$ simultaneously, which has obvious advantages if data collection is slow.

Perhaps most importantly, our algorithm also provides region estimates of the unknown parameters of the Hamiltonian.  This is to say that at the
end of the experiments we not only have an estimate of not only the values of the unknown parameters but also our uncertainty in those values.
We showed that the method is capable of providing region estimates that contain the actual unknown parameters with probability approximately $99.7\%$.  
Our algorithm also was used to provide construct similar confidence intervals for cases where the unknown Hamiltonian parameters were not constant
between experiments but instead fluctuated according to an unknown distribution.  These tests showed that our algorithm is robust and also is valuable
even if the cost of performing experiments is minimal.

Finally, we compared the costs in terms of experimental and computational time of our algorithm.  We found that the heuristics that reduced the experimental time most significantly often required more computational time to reach a desired level of accuracy.  Conversely, we found that many of the computationally innexpensive heuristics failed to reduce the mean--square errors after a finite number of experiments.  This suggests that the relative merits of different heuristics change as the relative costs of computational time and experimental time and the precision with which the unknown parameters should be estimated vary.  

An obvious extension of our work would be to consider more advanced optimization heuristics than conjugate gradient searches (such as particle swarm optimization algorithms).  Similarly, more advanced resampling techniques may lead to substantial reductions in the number of particles which in turn would reduce the computational cost of the algorithm.  Finally, estimates of how the number of experiments required to achieve a specific mean--square error scales with the number of unknown parameters would be an important extension of this work since it would assess the viability of these techniques for controlling and characterizing larger quantum systems.

\begin{acknowledgements}
This work was financially supported by the Canadian government through NSERC and CERC and by the United States government through DARPA.  NW would like to acknowledge funding from USARO-DTO.
\end{acknowledgements}


\begin{thebibliography}{49}
\expandafter\ifx\csname natexlab\endcsname\relax\def\natexlab#1{#1}\fi
\expandafter\ifx\csname bibnamefont\endcsname\relax
  \def\bibnamefont#1{#1}\fi
\expandafter\ifx\csname bibfnamefont\endcsname\relax
  \def\bibfnamefont#1{#1}\fi
\expandafter\ifx\csname citenamefont\endcsname\relax
  \def\citenamefont#1{#1}\fi
\expandafter\ifx\csname url\endcsname\relax
  \def\url#1{\texttt{#1}}\fi
\expandafter\ifx\csname urlprefix\endcsname\relax\def\urlprefix{URL }\fi
\providecommand{\bibinfo}[2]{#2}
\providecommand{\eprint}[2][]{\url{#2}}

\bibitem[{\citenamefont{Paris and Rehacek}(2004)}]{2004Quantum}
\bibinfo{editor}{\bibfnamefont{M.}~\bibnamefont{Paris}} \bibnamefont{and}
  \bibinfo{editor}{\bibfnamefont{J.}~\bibnamefont{Rehacek}}, eds.,
  \emph{\bibinfo{title}{{Quantum State Estimation}}}, vol.
  \bibinfo{volume}{649} of \emph{\bibinfo{series}{Lecture Notes in Physics}}
  (\bibinfo{publisher}{Springer}, \bibinfo{year}{2004}), ISBN
  \bibinfo{isbn}{978-3-540-22329-0},
  \urlprefix\url{http://www.springer.com/physics/quantum+physics/book/978-3-540-22329-0}.

\bibitem[{\citenamefont{Lanyon et~al.}(2011)\citenamefont{Lanyon, Hempel, Nigg,
  M\"uller, Gerritsma, Zähringer, Schindler, Barreiro, Rambach, Kirchmair
  et~al.}}]{LHM+11}
\bibinfo{author}{\bibfnamefont{B.~P.} \bibnamefont{Lanyon}},
  \bibinfo{author}{\bibfnamefont{C.}~\bibnamefont{Hempel}},
  \bibinfo{author}{\bibfnamefont{D.}~\bibnamefont{Nigg}},
  \bibinfo{author}{\bibfnamefont{M.}~\bibnamefont{M\"uller}},
  \bibinfo{author}{\bibfnamefont{R.}~\bibnamefont{Gerritsma}},
  \bibinfo{author}{\bibfnamefont{F.}~\bibnamefont{Zähringer}},
  \bibinfo{author}{\bibfnamefont{P.}~\bibnamefont{Schindler}},
  \bibinfo{author}{\bibfnamefont{J.~T.} \bibnamefont{Barreiro}},
  \bibinfo{author}{\bibfnamefont{M.}~\bibnamefont{Rambach}},
  \bibinfo{author}{\bibfnamefont{G.}~\bibnamefont{Kirchmair}},
  \bibnamefont{et~al.}, \bibinfo{journal}{Science}
  \textbf{\bibinfo{volume}{334}}, \bibinfo{pages}{57} (\bibinfo{year}{2011}),
  \eprint{http://www.sciencemag.org/content/334/6052/57.full.pdf},
  \urlprefix\url{http://www.sciencemag.org/content/334/6052/57.abstract}.

\bibitem[{\citenamefont{Gerritsma et~al.}(2011)\citenamefont{Gerritsma, Lanyon,
  Kirchmair, Z\"ahringer, Hempel, Casanova, Garcia-Ripoll, Solano, Blatt, and
  Roos}}]{GLK+11}
\bibinfo{author}{\bibfnamefont{R.}~\bibnamefont{Gerritsma}},
  \bibinfo{author}{\bibfnamefont{B.~P.} \bibnamefont{Lanyon}},
  \bibinfo{author}{\bibfnamefont{G.}~\bibnamefont{Kirchmair}},
  \bibinfo{author}{\bibfnamefont{F.}~\bibnamefont{Z\"ahringer}},
  \bibinfo{author}{\bibfnamefont{C.}~\bibnamefont{Hempel}},
  \bibinfo{author}{\bibfnamefont{J.}~\bibnamefont{Casanova}},
  \bibinfo{author}{\bibfnamefont{J.~J.} \bibnamefont{Garcia-Ripoll}},
  \bibinfo{author}{\bibfnamefont{E.}~\bibnamefont{Solano}},
  \bibinfo{author}{\bibfnamefont{R.}~\bibnamefont{Blatt}}, \bibnamefont{and}
  \bibinfo{author}{\bibfnamefont{C.~F.} \bibnamefont{Roos}},
  \bibinfo{journal}{Phys. Rev. Lett.} \textbf{\bibinfo{volume}{106}},
  \bibinfo{pages}{060503} (\bibinfo{year}{2011}),
  \urlprefix\url{http://link.aps.org/doi/10.1103/PhysRevLett.106.060503}.

\bibitem[{\citenamefont{{Kim} et~al.}(2010)\citenamefont{{Kim}, {Chang},
  {Korenblit}, {Islam}, {Edwards}, {Freericks}, {Lin}, {Duan}, and
  {Monroe}}}]{KCK+10}
\bibinfo{author}{\bibfnamefont{K.}~\bibnamefont{{Kim}}},
  \bibinfo{author}{\bibfnamefont{M.-S.} \bibnamefont{{Chang}}},
  \bibinfo{author}{\bibfnamefont{S.}~\bibnamefont{{Korenblit}}},
  \bibinfo{author}{\bibfnamefont{R.}~\bibnamefont{{Islam}}},
  \bibinfo{author}{\bibfnamefont{E.~E.} \bibnamefont{{Edwards}}},
  \bibinfo{author}{\bibfnamefont{J.~K.} \bibnamefont{{Freericks}}},
  \bibinfo{author}{\bibfnamefont{G.-D.} \bibnamefont{{Lin}}},
  \bibinfo{author}{\bibfnamefont{L.-M.} \bibnamefont{{Duan}}},
  \bibnamefont{and} \bibinfo{author}{\bibfnamefont{C.}~\bibnamefont{{Monroe}}},
  \bibinfo{journal}{\nat} \textbf{\bibinfo{volume}{465}}, \bibinfo{pages}{590}
  (\bibinfo{year}{2010}).

\bibitem[{\citenamefont{Bendersky et~al.}(2008)\citenamefont{Bendersky,
  Pastawski, and Paz}}]{BPP08}
\bibinfo{author}{\bibfnamefont{A.}~\bibnamefont{Bendersky}},
  \bibinfo{author}{\bibfnamefont{F.}~\bibnamefont{Pastawski}},
  \bibnamefont{and} \bibinfo{author}{\bibfnamefont{J.~P.} \bibnamefont{Paz}},
  \bibinfo{journal}{Phys. Rev. Lett.} \textbf{\bibinfo{volume}{100}},
  \bibinfo{pages}{190403} (\bibinfo{year}{2008}),
  \urlprefix\url{http://link.aps.org/doi/10.1103/PhysRevLett.100.190403}.

\bibitem[{\citenamefont{Bendersky et~al.}(2009)\citenamefont{Bendersky,
  Pastawski, and Paz}}]{BPP09}
\bibinfo{author}{\bibfnamefont{A.}~\bibnamefont{Bendersky}},
  \bibinfo{author}{\bibfnamefont{F.}~\bibnamefont{Pastawski}},
  \bibnamefont{and} \bibinfo{author}{\bibfnamefont{J.~P.} \bibnamefont{Paz}},
  \bibinfo{journal}{Phys. Rev. A} \textbf{\bibinfo{volume}{80}},
  \bibinfo{pages}{032116} (\bibinfo{year}{2009}),
  \urlprefix\url{http://link.aps.org/doi/10.1103/PhysRevA.80.032116}.

\bibitem[{\citenamefont{Mohseni and Rezakhani}(2009)}]{MR09}
\bibinfo{author}{\bibfnamefont{M.}~\bibnamefont{Mohseni}} \bibnamefont{and}
  \bibinfo{author}{\bibfnamefont{A.~T.} \bibnamefont{Rezakhani}},
  \bibinfo{journal}{Phys. Rev. A} \textbf{\bibinfo{volume}{80}},
  \bibinfo{pages}{010101} (\bibinfo{year}{2009}),
  \urlprefix\url{http://link.aps.org/doi/10.1103/PhysRevA.80.010101}.

\bibitem[{\citenamefont{Branderhorst et~al.}(2009)\citenamefont{Branderhorst,
  Nunn, Walmsley, and Kosut}}]{BNW+09}
\bibinfo{author}{\bibfnamefont{M.~P.~A.} \bibnamefont{Branderhorst}},
  \bibinfo{author}{\bibfnamefont{J.}~\bibnamefont{Nunn}},
  \bibinfo{author}{\bibfnamefont{I.~A.} \bibnamefont{Walmsley}},
  \bibnamefont{and} \bibinfo{author}{\bibfnamefont{R.~L.} \bibnamefont{Kosut}},
  \bibinfo{journal}{New Journal of Physics} \textbf{\bibinfo{volume}{11}},
  \bibinfo{pages}{115010} (\bibinfo{year}{2009}),
  \urlprefix\url{http://stacks.iop.org/1367-2630/11/i=11/a=115010}.

\bibitem[{\citenamefont{Flammia and Liu}(2011)}]{Flammia2011Direct}
\bibinfo{author}{\bibfnamefont{S.~T.} \bibnamefont{Flammia}} \bibnamefont{and}
  \bibinfo{author}{\bibfnamefont{Y.~K.} \bibnamefont{Liu}},
  \bibinfo{journal}{Physical Review Letters} \textbf{\bibinfo{volume}{106}},
  \bibinfo{pages}{230501+} (\bibinfo{year}{2011}),
  \urlprefix\url{http://dx.doi.org/10.1103/PhysRevLett.106.230501}.

\bibitem[{\citenamefont{da~Silva et~al.}(2011)\citenamefont{da~Silva, Cardinal,
  and Poulin}}]{daSilva2011Practical}
\bibinfo{author}{\bibfnamefont{M.~P.} \bibnamefont{da~Silva}},
  \bibinfo{author}{\bibfnamefont{O.~L.} \bibnamefont{Cardinal}},
  \bibnamefont{and} \bibinfo{author}{\bibfnamefont{D.}~\bibnamefont{Poulin}},
  \bibinfo{journal}{Physical Review Letters} \textbf{\bibinfo{volume}{107}},
  \bibinfo{pages}{210404+} (\bibinfo{year}{2011}),
  \urlprefix\url{http://dx.doi.org/10.1103/PhysRevLett.107.210404}.

\bibitem[{\citenamefont{Oi and Schirmer}(2012)}]{Oi2012Quantum}
\bibinfo{author}{\bibfnamefont{D.}~\bibnamefont{Oi}} \bibnamefont{and}
  \bibinfo{author}{\bibfnamefont{S.}~\bibnamefont{Schirmer}}
  (\bibinfo{year}{2012}), \eprint{1202.5779},
  \urlprefix\url{http://arxiv.org/abs/1202.5779}.

\bibitem[{\citenamefont{Doucet and Johansen}(2009)}]{Doucet2009Tutorial}
\bibinfo{author}{\bibfnamefont{A.}~\bibnamefont{Doucet}} \bibnamefont{and}
  \bibinfo{author}{\bibfnamefont{A.~M.} \bibnamefont{Johansen}},
  \emph{\bibinfo{title}{{A Tutorial on Particle Filtering and Smoothing:
  Fifteen Years Later}}} (\bibinfo{publisher}{Oxford University Press},
  \bibinfo{year}{2009}),
  \urlprefix\url{http://www.cs.ubc.ca/\~arnaud/doucet\_johansen\_tutorialPF.pdf}.

\bibitem[{\citenamefont{Loredo}(2004)}]{Loredo2004Bayesian}
\bibinfo{author}{\bibfnamefont{T.~J.} \bibnamefont{Loredo}},
  \bibinfo{journal}{AIP Conference Proceedings} \textbf{\bibinfo{volume}{707}},
  \bibinfo{pages}{330} (\bibinfo{year}{2004}),
  \urlprefix\url{http://dx.doi.org/10.1063/1.1751377}.

\bibitem[{\citenamefont{Kuck et~al.}(2006)\citenamefont{Kuck, de~Freitas, and
  Doucet}}]{Kuck2006SMC}
\bibinfo{author}{\bibfnamefont{H.}~\bibnamefont{Kuck}},
  \bibinfo{author}{\bibfnamefont{N.}~\bibnamefont{de~Freitas}},
  \bibnamefont{and} \bibinfo{author}{\bibfnamefont{A.}~\bibnamefont{Doucet}},
  in \emph{\bibinfo{booktitle}{Nonlinear Statistical Signal Processing
  Workshop, 2006 IEEE}} (\bibinfo{publisher}{IEEE}, \bibinfo{year}{2006}), pp.
  \bibinfo{pages}{99--102}, ISBN \bibinfo{isbn}{978-1-4244-0581-7},
  \urlprefix\url{http://dx.doi.org/10.1109/NSSPW.2006.4378829}.

\bibitem[{\citenamefont{Scarpa and Dunson}(2007)}]{Scarpa2007Bayesian}
\bibinfo{author}{\bibfnamefont{B.}~\bibnamefont{Scarpa}} \bibnamefont{and}
  \bibinfo{author}{\bibfnamefont{D.~B.} \bibnamefont{Dunson}},
  \bibinfo{journal}{Statist. Med.} \textbf{\bibinfo{volume}{26}},
  \bibinfo{pages}{1920} (\bibinfo{year}{2007}),
  \urlprefix\url{http://dx.doi.org/10.1002/sim.2846}.

\bibitem[{\citenamefont{Cavagnaro et~al.}(2010)\citenamefont{Cavagnaro, Pitt,
  and Myung}}]{Cavagnaro2010Adaptive}
\bibinfo{author}{\bibfnamefont{D.~R.} \bibnamefont{Cavagnaro}},
  \bibinfo{author}{\bibfnamefont{M.~A.} \bibnamefont{Pitt}}, \bibnamefont{and}
  \bibinfo{author}{\bibfnamefont{J.~I.} \bibnamefont{Myung}},
  \bibinfo{journal}{Advances in Neural Information Processing Systems}
  \textbf{\bibinfo{volume}{22}}, \bibinfo{pages}{234--242} (\bibinfo{year}{2010}).

\bibitem[{\citenamefont{Kantas et~al.}(2010)\citenamefont{Kantas,
  Lecchini-Visintini, and Maciejowski}}]{Kantas2010Simulationbased}
\bibinfo{author}{\bibfnamefont{N.}~\bibnamefont{Kantas}},
  \bibinfo{author}{\bibfnamefont{A.}~\bibnamefont{Lecchini-Visintini}},
  \bibnamefont{and} \bibinfo{author}{\bibfnamefont{J.~M.}
  \bibnamefont{Maciejowski}}, \bibinfo{journal}{Int. J. Adapt. Control Signal
  Process.} \textbf{\bibinfo{volume}{24}}, \bibinfo{pages}{882}
  (\bibinfo{year}{2010}), \urlprefix\url{http://dx.doi.org/10.1002/acs.1204}.

\bibitem[{\citenamefont{Huan and Marzouk}(2011)}]{Huan2011Simulationbased}
\bibinfo{author}{\bibfnamefont{X.}~\bibnamefont{Huan}} \bibnamefont{and}
  \bibinfo{author}{\bibfnamefont{Y.~M.} \bibnamefont{Marzouk}}
  (\bibinfo{year}{2011}), \eprint{1108.4146},
  \urlprefix\url{http://arxiv.org/abs/1108.4146}.

\bibitem[{\citenamefont{Husz\'{a}r and Houlsby}(2012)}]{Huszar2012Adaptive}
\bibinfo{author}{\bibfnamefont{F.}~\bibnamefont{Husz\'{a}r}} \bibnamefont{and}
  \bibinfo{author}{\bibfnamefont{N.~M.~T.} \bibnamefont{Houlsby}},
  \bibinfo{journal}{Physical Review A} \textbf{\bibinfo{volume}{85}}
  \bibinfo{pages}{052120}
  (\bibinfo{year}{2012}), ISSN \bibinfo{issn}{1094-1622}, \eprint{1107.0895},
  \urlprefix\url{http://dx.doi.org/10.1103/PhysRevA.85.052120}.

\bibitem[{\citenamefont{Bagan et~al.}(2006)\citenamefont{Bagan, Ballester,
  Gill, noz Tapia, and Isart}}]{Bagan2006Separable}
\bibinfo{author}{\bibfnamefont{E.}~\bibnamefont{Bagan}},
  \bibinfo{author}{\bibfnamefont{M.~A.} \bibnamefont{Ballester}},
  \bibinfo{author}{\bibfnamefont{R.~D.} \bibnamefont{Gill}},
  \bibinfo{author}{\bibfnamefont{M.}~\bibnamefont{noz Tapia}},
  \bibnamefont{and} \bibinfo{author}{\bibfnamefont{O.~R.} \bibnamefont{Isart}},
  \bibinfo{journal}{Physical Review Letters} \textbf{\bibinfo{volume}{97}},
  \bibinfo{pages}{130501+} (\bibinfo{year}{2006}),
  \urlprefix\url{http://dx.doi.org/10.1103/PhysRevLett.97.130501}.

\bibitem[{\citenamefont{Servedio and Gortler}(2004)}]{Servedio2004Equivalences}
\bibinfo{author}{\bibfnamefont{R.~A.} \bibnamefont{Servedio}} \bibnamefont{and}
  \bibinfo{author}{\bibfnamefont{S.~J.} \bibnamefont{Gortler}},
  \bibinfo{journal}{SIAM Journal on Computing} \textbf{\bibinfo{volume}{33}},
  \bibinfo{pages}{1067} (\bibinfo{year}{2004}), ISSN \bibinfo{issn}{0097-5397},
  \urlprefix\url{http://dx.doi.org/10.1137/S0097539704412910}.

\bibitem[{\citenamefont{A\"{i}meur et~al.}(2006)\citenamefont{A\"{i}meur,
  Brassard, and Gambs}}]{Aimeur2006Machine}
\bibinfo{author}{\bibfnamefont{E.}~\bibnamefont{A\"{i}meur}},
  \bibinfo{author}{\bibfnamefont{G.}~\bibnamefont{Brassard}}, \bibnamefont{and}
  \bibinfo{author}{\bibfnamefont{S.}~\bibnamefont{Gambs}}
  (\bibinfo{publisher}{Springer Berlin / Heidelberg}, \bibinfo{address}{Berlin,
  Heidelberg}, \bibinfo{year}{2006}), vol. \bibinfo{volume}{4013} of
  \emph{\bibinfo{series}{Lecture Notes in Computer Science}},
  chap.~\bibinfo{chapter}{37}, pp. \bibinfo{pages}{431--442}, ISBN
  \bibinfo{isbn}{978-3-540-34628-9},
  \urlprefix\url{http://dx.doi.org/10.1007/11766247\_37}.

\bibitem[{\citenamefont{Aaronson}(2007)}]{Aaronson2007Learnability}
\bibinfo{author}{\bibfnamefont{S.}~\bibnamefont{Aaronson}},
  \bibinfo{journal}{Proceedings of the Royal Society A: Mathematical, Physical
  and Engineering Science} \textbf{\bibinfo{volume}{463}},
  \bibinfo{pages}{3089--3114} (\bibinfo{year}{2007}), ISSN \bibinfo{issn}{1471-2946},
  \urlprefix\url{http://dx.doi.org/10.1098/rspa.2007.0113}.

\bibitem[{\citenamefont{Hentschel and Sanders}(2010)}]{Hentschel2010Machine}
\bibinfo{author}{\bibfnamefont{A.}~\bibnamefont{Hentschel}} \bibnamefont{and}
  \bibinfo{author}{\bibfnamefont{B.~C.} \bibnamefont{Sanders}},
  \bibinfo{journal}{Physical Review Letters} \textbf{\bibinfo{volume}{104}},
  \bibinfo{pages}{063603+} (\bibinfo{year}{2010}),
  \urlprefix\url{http://dx.doi.org/10.1103/PhysRevLett.104.063603}.

\bibitem[{\citenamefont{Pudenz and Lidar}(2011)}]{Pudenz2011Quantum}
\bibinfo{author}{\bibfnamefont{K.~L.} \bibnamefont{Pudenz}} \bibnamefont{and}
  \bibinfo{author}{\bibfnamefont{D.~A.} \bibnamefont{Lidar}}
  (\bibinfo{year}{2011}), \eprint{1109.0325},
  \urlprefix\url{http://arxiv.org/abs/1109.0325}.

\bibitem[{\citenamefont{Hentschel and Sanders}(2011)}]{Hentschel2011Efficient}
\bibinfo{author}{\bibfnamefont{A.}~\bibnamefont{Hentschel}} \bibnamefont{and}
  \bibinfo{author}{\bibfnamefont{B.~C.} \bibnamefont{Sanders}},
  \bibinfo{journal}{Physical Review Letters} \textbf{\bibinfo{volume}{107}},
  \bibinfo{pages}{233601} (\bibinfo{year}{2011}),
  \urlprefix\url{http://dx.doi.org/10.1103/PhysRevLett.107.233601}.

\bibitem[{\citenamefont{Sergeevich and
  Bartlett}(2012)}]{Sergeevich2012Optimizing}
\bibinfo{author}{\bibfnamefont{A.}~\bibnamefont{Sergeevich}} \bibnamefont{and}
  \bibinfo{author}{\bibfnamefont{S.~D.} \bibnamefont{Bartlett}}
  (\bibinfo{year}{2012}), \eprint{1206.3830},
  \urlprefix\url{http://arxiv.org/abs/1206.3830}.
  
\bibitem[{\citenamefont{Caves}(1986)}]{caves_quantum_1986}
\bibinfo{author}{\bibfnamefont{C.~M.} \bibnamefont{Caves}},
  \bibinfo{journal}{Physical Review D} \textbf{\bibinfo{volume}{33}},
  \bibinfo{pages}{1643} (\bibinfo{year}{1986}),
  \urlprefix\url{http://link.aps.org/doi/10.1103/PhysRevD.33.1643}.

\bibitem[{\citenamefont{Tsang}(2009)}]{Tsang2009TimeSymmetric}
\bibinfo{author}{\bibfnamefont{M.}~\bibnamefont{Tsang}},
  \bibinfo{journal}{Physical Review Letters} \textbf{\bibinfo{volume}{102}},
  \bibinfo{pages}{250403+} (\bibinfo{year}{2009}),
  \urlprefix\url{http://dx.doi.org/10.1103/PhysRevLett.102.250403}.

\bibitem[{\citenamefont{Wheatley et~al.}(2010)\citenamefont{Wheatley, Berry,
  Yonezawa, Nakane, Arao, Pope, Ralph, Wiseman, Furusawa, and
  Huntington}}]{Wheatley2010Adaptive}
\bibinfo{author}{\bibfnamefont{T.~A.} \bibnamefont{Wheatley}},
  \bibinfo{author}{\bibfnamefont{D.~W.} \bibnamefont{Berry}},
  \bibinfo{author}{\bibfnamefont{H.}~\bibnamefont{Yonezawa}},
  \bibinfo{author}{\bibfnamefont{D.}~\bibnamefont{Nakane}},
  \bibinfo{author}{\bibfnamefont{H.}~\bibnamefont{Arao}},
  \bibinfo{author}{\bibfnamefont{D.~T.} \bibnamefont{Pope}},
  \bibinfo{author}{\bibfnamefont{T.~C.} \bibnamefont{Ralph}},
  \bibinfo{author}{\bibfnamefont{H.~M.} \bibnamefont{Wiseman}},
  \bibinfo{author}{\bibfnamefont{A.}~\bibnamefont{Furusawa}}, \bibnamefont{and}
  \bibinfo{author}{\bibfnamefont{E.~H.} \bibnamefont{Huntington}},
  \bibinfo{journal}{Physical Review Letters} \textbf{\bibinfo{volume}{104}},
  \bibinfo{pages}{093601+} (\bibinfo{year}{2010}),
  \urlprefix\url{http://dx.doi.org/10.1103/PhysRevLett.104.093601}.

\bibitem[{\citenamefont{Lindley}(1956)}]{Lindley1956On}
\bibinfo{author}{\bibfnamefont{D.~V.} \bibnamefont{Lindley}},
  \bibinfo{journal}{The Annals of Mathematical Statistics}
  \textbf{\bibinfo{volume}{27}}, \bibinfo{pages}{986} (\bibinfo{year}{1956}),
  ISSN \bibinfo{issn}{0003-4851},
  \urlprefix\url{http://dx.doi.org/10.1214/aoms/1177728069}.

\bibitem[{\citenamefont{Said et~al.}(2011)\citenamefont{Said, Berry, and
  Twamley}}]{SBT11}
\bibinfo{author}{\bibfnamefont{R.~S.} \bibnamefont{Said}},
  \bibinfo{author}{\bibfnamefont{D.~W.} \bibnamefont{Berry}}, \bibnamefont{and}
  \bibinfo{author}{\bibfnamefont{J.}~\bibnamefont{Twamley}},
  \bibinfo{journal}{Phys. Rev. B} \textbf{\bibinfo{volume}{83}},
  \bibinfo{pages}{125410} (\bibinfo{year}{2011}),
  \urlprefix\url{http://link.aps.org/doi/10.1103/PhysRevB.83.125410}.

\bibitem[{\citenamefont{Lehmann and Casella}(1998)}]{Lehmann1998Theory}
\bibinfo{author}{\bibfnamefont{E.~L.} \bibnamefont{Lehmann}} \bibnamefont{and}
  \bibinfo{author}{\bibfnamefont{G.}~\bibnamefont{Casella}},
  \emph{\bibinfo{title}{{Theory of Point Estimation}}}
  (\bibinfo{publisher}{Springer}, \bibinfo{year}{1998}), \bibinfo{edition}{2nd}
  ed., ISBN \bibinfo{isbn}{0387985026},
  \urlprefix\url{http://www.amazon.com/exec/obidos/redirect?tag=citeulike07-20\&path=ASIN/0387985026}.

\bibitem[{\citenamefont{Berger}(1985)}]{Berger1985Statistical}
\bibinfo{author}{\bibfnamefont{J.~O.} \bibnamefont{Berger}},
  \emph{\bibinfo{title}{{Statistical Decision Theory and Bayesian Analysis}}}
  (\bibinfo{publisher}{Springer}, \bibinfo{year}{1985}), \bibinfo{edition}{2nd}
  ed., ISBN \bibinfo{isbn}{0387960988},
  \urlprefix\url{http://www.amazon.com/exec/obidos/redirect?tag=citeulike07-20\&path=ASIN/0387960988}.

\bibitem[{\citenamefont{Blume-Kohout and
  Hayden}(2006)}]{BlumeKohout2006Accurate}
\bibinfo{author}{\bibfnamefont{R.}~\bibnamefont{Blume-Kohout}}
  \bibnamefont{and} \bibinfo{author}{\bibfnamefont{P.}~\bibnamefont{Hayden}}
  (\bibinfo{year}{2006}), \eprint{quant-ph/0603116},
  \urlprefix\url{http://arxiv.org/abs/quant-ph/0603116}.

\bibitem[{\citenamefont{Gill and Levit}(1995)}]{Gill1995Applications}
\bibinfo{author}{\bibfnamefont{R.~D.} \bibnamefont{Gill}} \bibnamefont{and}
  \bibinfo{author}{\bibfnamefont{B.~Y.} \bibnamefont{Levit}},
  \bibinfo{journal}{Bernoulli} \textbf{\bibinfo{volume}{1}} \bibinfo{pages}{59}
  (\bibinfo{year}{1995}), ISSN \bibinfo{issn}{13507265},
  \urlprefix\url{http://dx.doi.org/10.2307/3318681}.

\bibitem[{\citenamefont{Tichavsky et~al.}(1998)\citenamefont{Tichavsky,
  Muravchik, and Nehorai}}]{Tichavsky1998Posterior}
\bibinfo{author}{\bibfnamefont{P.}~\bibnamefont{Tichavsky}},
  \bibinfo{author}{\bibfnamefont{C.~H.} \bibnamefont{Muravchik}},
  \bibnamefont{and} \bibinfo{author}{\bibfnamefont{A.}~\bibnamefont{Nehorai}},
  \bibinfo{journal}{Signal Processing, IEEE Transactions on}
  \textbf{\bibinfo{volume}{46}}, \bibinfo{pages}{1386} (\bibinfo{year}{1998}),
  ISSN \bibinfo{issn}{1053-587X},
  \urlprefix\url{http://dx.doi.org/10.1109/78.668800}.

\bibitem[{\citenamefont{Gordon et~al.}(1993)\citenamefont{Gordon, Salmond, and
  Smith}}]{Gordon1993Novel}
\bibinfo{author}{\bibfnamefont{N.~J.} \bibnamefont{Gordon}},
  \bibinfo{author}{\bibfnamefont{D.~J.} \bibnamefont{Salmond}},
  \bibnamefont{and} \bibinfo{author}{\bibfnamefont{A.~F.~M.}
  \bibnamefont{Smith}}, \bibinfo{journal}{Radar and Signal Processing, IEE
  Proceedings F} \textbf{\bibinfo{volume}{140}}, \bibinfo{pages}{107--113}
  (\bibinfo{year}{1993}), ISSN \bibinfo{issn}{0956-375X},
  \urlprefix\url{http://dx.doi.org/10.1049/ip-f-2.1993.0015}.

\bibitem[{\citenamefont{Liu and West}(2000)}]{Liu2000Combined}
\bibinfo{author}{\bibfnamefont{J.}~\bibnamefont{Liu}} \bibnamefont{and}
  \bibinfo{author}{\bibfnamefont{M.}~\bibnamefont{West}},
  \emph{\bibinfo{title}{{Combined parameter and state estimation in
  simulation-based filtering}}} (\bibinfo{publisher}{Springer-Verlag},
  \bibinfo{year}{2000}).

\bibitem[{\citenamefont{Edwards et~al.}(1963)\citenamefont{Edwards, Lindman,
  and Savage}}]{edwards_bayesian_1963}
\bibinfo{author}{\bibfnamefont{W.}~\bibnamefont{Edwards}},
  \bibinfo{author}{\bibfnamefont{H.}~\bibnamefont{Lindman}}, \bibnamefont{and}
  \bibinfo{author}{\bibfnamefont{L.~J.} \bibnamefont{Savage}},
  \bibinfo{journal}{Psychological Review} \textbf{\bibinfo{volume}{70}},
  \bibinfo{pages}{193} (\bibinfo{year}{1963}), ISSN
  \bibinfo{issn}{1939-{1471(Electronic);0033-295X(Print)}}.
  
\bibitem[{\citenamefont{Bernardo}(2005)}]{bernardo_intrinsic_2005}
\bibinfo{author}{\bibfnamefont{J.}~\bibnamefont{Bernardo}},
  \bibinfo{journal}{{TEST}} \textbf{\bibinfo{volume}{14}}, \bibinfo{pages}{317}
  (\bibinfo{year}{2005}), ISSN \bibinfo{issn}{1133-0686},
  \urlprefix\url{http://www.springerlink.com/content/u823757117165122/abstract/}.
  
\bibitem[{\citenamefont{Wackerly et~al.}(2001)\citenamefont{Wackerly,
  Mendenhall, and Scheaffer}}]{wackerly_mathematical_2001}
\bibinfo{author}{\bibfnamefont{D.}~\bibnamefont{Wackerly}},
  \bibinfo{author}{\bibfnamefont{W.}~\bibnamefont{Mendenhall}},
  \bibnamefont{and} \bibinfo{author}{\bibfnamefont{R.~L.}
  \bibnamefont{Scheaffer}}, \emph{\bibinfo{title}{Mathematical Statistics with
  Applications {(Mathematical} Statistics}} (\bibinfo{publisher}{Duxbury
  Press}, \bibinfo{year}{2001}), \bibinfo{edition}{6th} ed., ISBN
  \bibinfo{isbn}{0534377416}.
  
\bibitem[{\citenamefont{Beale}(1960)}]{beale_confidence_1960}
\bibinfo{author}{\bibfnamefont{E.~M.~L.} \bibnamefont{Beale}},
  \bibinfo{journal}{Journal of the Royal Statistical Society. Series B
  {(Methodological)}} \textbf{\bibinfo{volume}{22}}, \bibinfo{pages}{41}
  (\bibinfo{year}{1960}), ISSN \bibinfo{issn}{0035-9246},
  \bibinfo{note}{{ArticleType:} research-article / Full publication date: 1960
  / Copyright © 1960 Royal Statistical Society},
  \urlprefix\url{http://www.jstor.org/stable/2983877}.
  
\bibitem[{\citenamefont{Blume-Kohout}(2012)}]{BlumeKohout2012Robust}
\bibinfo{author}{\bibfnamefont{R.}~\bibnamefont{Blume-Kohout}},
  \emph{\bibinfo{title}{{Robust error bars for quantum tomography}}}
  (\bibinfo{year}{2012}), \eprint{1202.5270},
  \urlprefix\url{http://arxiv.org/abs/1202.5270}.

\bibitem[{\citenamefont{Christandl and Renner}(2011)}]{Christandl2011Reliable}
\bibinfo{author}{\bibfnamefont{M.}~\bibnamefont{Christandl}} \bibnamefont{and}
  \bibinfo{author}{\bibfnamefont{R.}~\bibnamefont{Renner}}
  (\bibinfo{year}{2011}), \eprint{1108.5329},
  \urlprefix\url{http://arxiv.org/abs/1108.5329}.
  
\bibitem[{\citenamefont{Todd and Yıldırım}(2007)}]{todd_khachiyans_2007}
\bibinfo{author}{\bibfnamefont{M.~J.} \bibnamefont{Todd}} \bibnamefont{and}
  \bibinfo{author}{\bibfnamefont{E.~A.} \bibnamefont{Yıldırım}},
  \bibinfo{journal}{Discrete Applied Mathematics}
  \textbf{\bibinfo{volume}{155}}, \bibinfo{pages}{1731} (\bibinfo{year}{2007}),
  ISSN \bibinfo{issn}{{0166-218X}},
  \urlprefix\url{http://www.sciencedirect.com/science/article/pii/S0166218X07000716}.

\bibitem[{\citenamefont{Barber et~al.}(1996)\citenamefont{Barber, Dobkin, and
  Huhdanpaa}}]{barber_quickhull_1996}
\bibinfo{author}{\bibfnamefont{C.~B.} \bibnamefont{Barber}},
  \bibinfo{author}{\bibfnamefont{D.~P.} \bibnamefont{Dobkin}},
  \bibnamefont{and}
  \bibinfo{author}{\bibfnamefont{H.}~\bibnamefont{Huhdanpaa}},
  \bibinfo{journal}{{ACM} Trans. Math. Softw.} \textbf{\bibinfo{volume}{22}},
  \bibinfo{pages}{469–483} (\bibinfo{year}{1996}), ISSN
  \bibinfo{issn}{0098-3500},
  \urlprefix\url{http://doi.acm.org/10.1145/235815.235821}.


\bibitem[{\citenamefont{Sergeevich et~al.}(2011)\citenamefont{Sergeevich,
  Chandran, Combes, Bartlett, and Wiseman}}]{Sergeevich2011Characterization}
\bibinfo{author}{\bibfnamefont{A.}~\bibnamefont{Sergeevich}},
  \bibinfo{author}{\bibfnamefont{A.}~\bibnamefont{Chandran}},
  \bibinfo{author}{\bibfnamefont{J.}~\bibnamefont{Combes}},
  \bibinfo{author}{\bibfnamefont{S.}~\bibnamefont{Bartlett}}, \bibnamefont{and}
  \bibinfo{author}{\bibfnamefont{H.}~\bibnamefont{Wiseman}},
  \bibinfo{journal}{Physical Review A} \textbf{\bibinfo{volume}{84}} \bibinfo{pages}{052315}
  (\bibinfo{year}{2011}), ISSN \bibinfo{issn}{1094-1622}, \eprint{1102.3700},
  \urlprefix\url{http://dx.doi.org/10.1103/PhysRevA.84.052315}.

\bibitem[{\citenamefont{Ferrie et~al.}(2012{\natexlab{a}})\citenamefont{Ferrie,
  Granade, and Cory}}]{ferrie_adaptive_2012}
\bibinfo{author}{\bibfnamefont{C.}~\bibnamefont{Ferrie}},
  \bibinfo{author}{\bibfnamefont{C.~E.} \bibnamefont{Granade}},
  \bibnamefont{and} \bibinfo{author}{\bibfnamefont{D.~G.} \bibnamefont{Cory}},
  \bibinfo{journal}{{AIP} Conference Proceedings}
  \textbf{\bibinfo{volume}{1443}}, \bibinfo{pages}{165}
  (\bibinfo{year}{2012}{\natexlab{a}}), ISSN \bibinfo{issn}{{0094243X}},
  \urlprefix\url{http://proceedings.aip.org/resource/2/apcpcs/1443/1/165_1?isAuthorized=no}.

\bibitem[{\citenamefont{Ferrie et~al.}(2012{\natexlab{b}})\citenamefont{Ferrie,
  Granade, and Cory}}]{ferrie_how_2012}
\bibinfo{author}{\bibfnamefont{C.}~\bibnamefont{Ferrie}},
  \bibinfo{author}{\bibfnamefont{C.}~\bibnamefont{Granade}}, \bibnamefont{and}
  \bibinfo{author}{\bibfnamefont{D.}~\bibnamefont{Cory}},
  \bibinfo{journal}{Quantum Information Processing} \textit{\bibinfo{volume}{Online First}} pp. \bibinfo{pages}{1--13}
  (\bibinfo{year}{2012}{\natexlab{b}}), ISSN \bibinfo{issn}{1570-0755},
  \urlprefix\url{http://www.springerlink.com/content/130hm02564t34j84/abstract/}.

\bibitem[{\citenamefont{Lindblad}(1976)}]{lindblad_generators_1976}
\bibinfo{author}{\bibfnamefont{G.}~\bibnamefont{Lindblad}},
  \bibinfo{journal}{Communications in Mathematical Physics (1965-1997)}
  \textbf{\bibinfo{volume}{48}}, \bibinfo{pages}{119} (\bibinfo{year}{1976}),
  ISSN \bibinfo{issn}{0010-3616, 1432-0916},
  \urlprefix\url{http://projecteuclid.org/euclid.cmp/1103899849}.

\bibitem[{imp()}]{imp_details}
\bibinfo{note}{For the numerical experiments, Algorithm \ref{alg:smc-complete}
  was implemented using version 0.9.0 of the SciPy package \cite{SciPy2001}
  with Python versions 2.7.1, 2.7.2 and 2.7.3.}

\bibitem[{\citenamefont{Boulant et~al.}(2003)\citenamefont{Boulant, Havel,
  Pravia, and Cory}}]{Boulant2003Robust}
\bibinfo{author}{\bibfnamefont{N.}~\bibnamefont{Boulant}},
  \bibinfo{author}{\bibfnamefont{T.~F.} \bibnamefont{Havel}},
  \bibinfo{author}{\bibfnamefont{M.~A.} \bibnamefont{Pravia}},
  \bibnamefont{and} \bibinfo{author}{\bibfnamefont{D.~G.} \bibnamefont{Cory}},
  \bibinfo{journal}{Physical Review A} \textbf{\bibinfo{volume}{67}},
  \bibinfo{pages}{042322+} (\bibinfo{year}{2003}),
  \urlprefix\url{http://dx.doi.org/10.1103/PhysRevA.67.042322}.

\bibitem[{\citenamefont{Weinstein et~al.}(2004)\citenamefont{Weinstein, Havel,
  Emerson, Boulant, Saraceno, Lloyd, and Cory}}]{Weinstein2004Quantum}
\bibinfo{author}{\bibfnamefont{Y.~S.} \bibnamefont{Weinstein}},
  \bibinfo{author}{\bibfnamefont{T.~F.} \bibnamefont{Havel}},
  \bibinfo{author}{\bibfnamefont{J.}~\bibnamefont{Emerson}},
  \bibinfo{author}{\bibfnamefont{N.}~\bibnamefont{Boulant}},
  \bibinfo{author}{\bibfnamefont{M.}~\bibnamefont{Saraceno}},
  \bibinfo{author}{\bibfnamefont{S.}~\bibnamefont{Lloyd}}, \bibnamefont{and}
  \bibinfo{author}{\bibfnamefont{D.~G.} \bibnamefont{Cory}},
  \bibinfo{journal}{The Journal of Chemical Physics}
  \textbf{\bibinfo{volume}{121}}, \bibinfo{pages}{6117} (\bibinfo{year}{2004}),
  \urlprefix\url{http://dx.doi.org/10.1063/1.1785151}.

\bibitem[{\citenamefont{Boulant et~al.}(2004)\citenamefont{Boulant, Emerson,
  Havel, Cory, and Furuta}}]{Boulant2004Incoherent}
\bibinfo{author}{\bibfnamefont{N.}~\bibnamefont{Boulant}},
  \bibinfo{author}{\bibfnamefont{J.}~\bibnamefont{Emerson}},
  \bibinfo{author}{\bibfnamefont{T.}~\bibnamefont{Havel}},
  \bibinfo{author}{\bibfnamefont{D.}~\bibnamefont{Cory}}, \bibnamefont{and}
  \bibinfo{author}{\bibfnamefont{S.}~\bibnamefont{Furuta}},
  \bibinfo{journal}{The Journal of Chemical Physics}
  \textbf{\bibinfo{volume}{121}}, \bibinfo{pages}{2955} (\bibinfo{year}{2004}),
  \urlprefix\url{http://dx.doi.org/10.1063/1.1773161}.

\bibitem[{\citenamefont{Gutierrez et~al.}(2008)\citenamefont{Gutierrez, Romero,
  Trenas, and Zapata}}]{gutierrez_parallel_2008}
\bibinfo{author}{\bibfnamefont{E.}~\bibnamefont{Gutierrez}},
  \bibinfo{author}{\bibfnamefont{S.}~\bibnamefont{Romero}},
  \bibinfo{author}{\bibfnamefont{M.}~\bibnamefont{Trenas}}, \bibnamefont{and}
  \bibinfo{author}{\bibfnamefont{E.}~\bibnamefont{Zapata}}, in
  \emph{\bibinfo{booktitle}{Computational Science – {ICCS} 2008}}, edited by
  \bibinfo{editor}{\bibfnamefont{M.}~\bibnamefont{Bubak}},
  \bibinfo{editor}{\bibfnamefont{G.}~\bibnamefont{van Albada}},
  \bibinfo{editor}{\bibfnamefont{J.}~\bibnamefont{Dongarra}}, \bibnamefont{and}
  \bibinfo{editor}{\bibfnamefont{P.}~\bibnamefont{Sloot}}
  (\bibinfo{publisher}{Springer Berlin / Heidelberg}, \bibinfo{year}{2008}),
  vol. \bibinfo{volume}{5101} of \emph{\bibinfo{series}{Lecture Notes in
  Computer Science}}, pp. \bibinfo{pages}{700--709}, ISBN
  \bibinfo{isbn}{978-3-540-69383-3},
  \urlprefix\url{http://dx.doi.org/10.1007/978-3-540-69384-0_75}.

\bibitem[{\citenamefont{Khalid et~al.}(2004)\citenamefont{Khalid, Zilic, and
  Radecka}}]{khalid_fpga_2004}
\bibinfo{author}{\bibfnamefont{A.}~\bibnamefont{Khalid}},
  \bibinfo{author}{\bibfnamefont{Z.}~\bibnamefont{Zilic}}, \bibnamefont{and}
  \bibinfo{author}{\bibfnamefont{K.}~\bibnamefont{Radecka}}, in
  \emph{\bibinfo{booktitle}{Computer Design: {VLSI} in Computers and
  Processors, 2004. {ICCD} 2004. Proceedings. {IEEE} International Conference
  on}} (\bibinfo{year}{2004}), pp. \bibinfo{pages}{310 -- 315},
  \urlprefix\url{http://dx.doi.org/10.1109/ICCD.2004.1347938}.

\bibitem[{\citenamefont{Jones et~al.}(2001--)\citenamefont{Jones, Oliphant,
  Peterson et~al.}}]{SciPy2001}
\bibinfo{author}{\bibfnamefont{E.}~\bibnamefont{Jones}},
  \bibinfo{author}{\bibfnamefont{T.}~\bibnamefont{Oliphant}},
  \bibinfo{author}{\bibfnamefont{P.}~\bibnamefont{Peterson}},
  \bibnamefont{et~al.}, \emph{\bibinfo{title}{{SciPy}: Open source scientific
  tools for {Python}}} (\bibinfo{year}{2001--}),
  \urlprefix\url{http://www.scipy.org/}.

\end{thebibliography}

\end{document}